\documentclass[aps,pre,onecolumn,floatfix]{revtex4}
\usepackage{graphicx}
\usepackage{amssymb,amsfonts,amsmath}
\usepackage{mathtools}
\usepackage{epsf}
\usepackage{subfigure}
\usepackage{epstopdf}

\newcommand{\be}{\begin{equation}}
\newcommand{\ee}{\end{equation}}
\newcommand{\bea}{\begin{eqnarray}}
\newcommand{\eea}{\end{eqnarray}}

\newcommand{\dd}{{\mathrm d}}

\begin{document}

\title{\bf Kinetics of template-directed multistate copolymerization}

\author{Pierre Gaspard}
\affiliation{Center for Nonlinear Phenomena and Complex Systems,\\
Universit\'e Libre de Bruxelles (ULB), Code Postal 231, Campus Plaine,
B-1050 Brussels, Belgium}

\begin{abstract}
We consider processes of template-directed multistate copolymerization by molecular machines such as polymerases or ribosomes, having multiple states of conformation or activation.  We show that the kinetic equations of these processes can be exactly solved for the mean growth velocity, the sequence probabilities of the grown copy, and the local probabilities and fractions of monomeric units in the copy.  Asymptotically, in the long-time limit, the kinetic equations are solved with a matrix factorization ansatz in terms of a backward iteration, forming an iterated matrix function system, and a complementary forward iteration, both running along the template sequence.  The iterative method is very significantly faster than usual computational methods, as demonstrated with a numerical example.
\end{abstract}

\noindent 
\vskip 0.5 cm

\maketitle

\section{Introduction}

In our cells, genetic information is encoded at the molecular level in the form of DNA, which is transcribed into RNA and then translated into proteins.  Beyond, it is transmitted from generation to generation upon DNA replication.  The molecules of DNA, RNA, and proteins are copolymers, i.e., polymers composed of several types of monomeric units: the deoxyribonucleotides $\{$A,C,G,T$\}$ for DNA, the nucleotides $\{$A,C,G,U$\}$ for RNA, and twenty amino acids for proteins.  In this regard, DNA replication, its transcription, and its translation are templated-directed copolymerization processes.  These processes are catalyzed by molecular machines, which are DNA polymerases for replication, RNA polymerases for transcription, and ribosomes for translation.  In general, these machines function by performing transitions between several conformational or activation states.

For replication, the work by Johnson and coworkers has shown that high-fidelity DNA polymerases like that of virus~T7 exhibit several distinct structural states, between which large changes in conformation occur, playing an important role in the discrimination between correct and incorrect base pairs \cite{TJ06,J10}.  These nucleotide-induced conformational changes have been detected with fluorescence methods \cite{DJ20,DJ21,DKJ22}.

For transcription, ARN polymerases are known to have pathways into backtracking and forward-tracking translocation modes in addition to their main reaction pathway between their pre- and post-translocation states \cite{BSW04,LFW07,DILKLKB13,IDCLLPKB14}.

For translation, ribosomes utilize transitions between several conformational and activation states.  A mechanochemical cycle of five states has been considered \cite{GCCR09} to model the experimental observations of translation by single ribosomes \cite{WLHZYNBT08}.  In addition, two proofreading backward transitions have been discovered among these states \cite{IUSE16,LDK20}.

Therefore, these molecular machines typically function with transitions occurring in a network of internal states between every elongation step during the synthesis of the copolymer and the motion of the machine along the template.  These internal states should thus be taken into account in the mathematical description of the kinetics of these machines, in addition to the heterogeneity of the template, which may form an arbitrary sequence of units containing genetic information.

These aspects go beyond the description simplifying templated-directed copolymerization into free copolymerization, where the copy is reduced to a sequence of correct or incorrect (right or wrong) units \cite{B79,AG09,GA14,SP15,G16a,G16b}.  For such free copolymerization processes, the exact asymptotic solution of the mathematical theory is given by a closed set of nonlinear equations, which has been confirmed in recent publications \cite{MKI24a,MKI24b,AMG26}.  Furthermore, a method based on the theory of Markov stochastic processes has been proposed to deal with templated copolymerization with an arbitrarily complex network of reversible reactions \cite{QJPBO23,GQO25}.  If the catalyst has only a single state, the heterogeneity of the template can be described exactly using iterated function systems \cite{BD85}, where a function is associated with each unit or codon of the template \cite{G16PRL,G17JSM,G17PRE}.  In this case, an equivalent theory based on a first-passage approach has been developed to study DNA polymerase fidelity \cite{LZSOL19,LSOL21}.

In the presence of multiple states of conformation or activation, the scheme should be generalized using a matrix formulation.  The matrices allow us to describe the transitions between the distinct states of the catalyst.  This approach was pioneered by Coleman and Fox for irreversible free polymerization processes \cite{CF63JCP,CF63JACS,CF63JPS} and shown to also give the exact asymptotic solution for reversible free and template-directed copolymerization processes \cite{G19JCP,G20}.

Here, our goal is to extend this approach to general multistate processes, in which all the transitions are possible from and to every internal states, and upon attachment or detachment of monomeric units to the copy sequence.  We start from the mathematical formulation in terms of the kinetic equations, i.e., the master equations ruling the time evolution for the probabilities of finding the system formed by the catalyst, the template, and its copy into any one of its particular states.  The master equations are linear and they can thus be solved using the methods of linear analysis.  The master equations are defined in terms of the rates of the stochastic process we consider, which is a Markov jump process \cite{vK81,Gardiner,S16}.  These rates depend on a certain number of monomeric units, which have already been incorporated at the growing tip of the copy, on the monomeric units of the template around the location of this tip, and on the internal states of the catalyst.  The rates may also depend on the concentrations of monomers and other molecules activating the process, these molecular species being found in the solution surrounding the catalyst.  Next, the master equations are solved with the methods of mathematical analysis to obtain, in particular, the mean growth velocity and the probabilities of possible sequences for the copy.  In order to solve the master equations, they are first rewritten in a more compact matrix form, dealing with the multiple states of the catalyst.  Next, a hierarchy of nested equations is deduced for the probabilities of the different possible subsequences at the growing tip of the copy.  As soon as the length of these subsequences is longer than the number of monomeric units of the copy, on which the rates depend, the equations of the hierarchy are found to all have a similar structure.  This key observation allows us to solve the hierarchy with a matrix factorization ansatz, as pioneered by Coleman and Fox \cite{CF63JCP,CF63JACS,CF63JPS}.  Asymptotically, in the long-time limit, this ansatz leads to an iterated matrix function system (IMFS) running backward along the template.  This backward iteration is combined with an associated forward iteration to obtain the exact solution for the mean growth velocity and the probabilities of the copy sequences.  This method also applies to the stationary regime where the growth of the copy is stalled.  The great advantage of the IMFS method is that it is more than a million times faster than the numerical simulation method based on Gillespie's algorithm, while always giving the exact asymptotic solution of the problem.

After this introduction, the paper is organized according to the following plan.  Section~\ref{sec:math} presents the mathematical formulation with the description of the stochastic process, the kinetic equations, the observable quantities, the different regimes of the process, and the matrix formulation.  Section~\ref{sec:sol} gives the solutions of the kinetic equations, starting from the deduction of the hierarchy of nested equations for the probabilities of subsequences of longer and longer lengths at the growing tip of the copy, continuing with the decomposition of these equations into particular solutions according to the principle of superposition for solving linear equations.  The matrix factorization ansatz is enunciated.  The solution of the mathematical problem is given in terms of the backward and forward iterations, leading to expressions for the mean growth velocity, the sequence probabilities, and the local probabilities and fractions of the different types of monomeric units in the copy.  The solution is also given in the stationary regime.  In Section~\ref{sec:num_example}, the method is applied to a numerical example to confirm its accuracy and to show its efficiency.  Section~\ref{sec:Conclusion} presents the conclusion and perspectives.  Appendix~\ref{AppA} gives the mathematical details leading to the exact asymptotic solution of the kinetic equations.  Appendix~\ref{AppB} is devoted to the intermediate regime of sublinear growth in time of the copy.  Appendix~\ref{AppC} summarizes the application of Gillespie's algorithm to the numerical example of Section~\ref{sec:num_example}.

\section{The mathematical framework}
\label{sec:math}

\subsection{The template-directed process}

We consider the catalytic formation of a copolymer chain made from monomeric units $m_l \in\{1,2,\dots,M\}$ and growing along a given semi-infinite template sequence $n_1n_2\cdots n_l n_{l+1}\cdots$ composed of units $n_l \in\{1,2,\dots,N\}$.  For instance, there are $M=N=4$ nucleotides in DNA replication and transcription, while $M=20$ amino acids and $N=4^3=64$ codons are involved in translation.

The template sequence may be homogeneous (i.e., composed of a single unit type with $n_l=n$ for all $l=1,2,3,\dots$), periodic with a prime period~$L$ (i.e., such that $n_{l+L}=n_l$ for all $l$), or disordered (i.e., with an infinite period, in which case the template is aperiodic).  In the latter case, the template sequence should be characterized by the probability distribution of its units.  The template is a Bernoulli chain if the distribution factorizes as $\nu(n_1n_2\cdots n_L)=\prod_{l=1}^{L} \nu(n_l)$ in terms of the unit fractions $\nu(n)$, or it may be a Markov chain of $k^{\rm th}$ order.  The template sequence remains invariant during the whole process.

The synthesis of the copolymer chain is promoted by a catalyst, which may be in either one or another of multiple states $i = 1,2,\dots,I$, such as the conformational states of the molecular machine performing replication, transcription, or translation.  If the process has some fidelity, the copolymer chain represents a copy of the template, which is imperfect due to possible errors in replication, transcription, or translation.

The synthesized copolymer chain $m_1m_2\cdots m_l$, its length $l$, and the state $i$ of the catalyst change in time according to the following reactions
\be
\begin{array}{l}
m_1m_2\cdots m_{l-1},\, i \qquad\qquad  + \ m_{l}\\
n_1\ n_2\; \cdots \, n_{l-1}\, n_{l} \ n_{l+1}\ \cdots
\end{array}
\qquad
\underset{w_{-m_l,l}^{j \to i}}{\overset{w_{+m_l,l}^{i \to j}}{\rightleftharpoons}}
\qquad
\begin{array}{l}
m_1m_2\cdots m_{l-1} m_{l},\, j\\
n_1\ n_2\; \cdots \, n_{l-1}\, n_{l} \ n_{l+1}\ \cdots
\end{array}
\qquad
(l\ge 1; i,j\in\{1,2,\dots,I\}) \, ,
\label{kin_1}
\ee
and
\be
\quad
\begin{array}{l}
m_1m_2\cdots m_{l-1} m_{l},\, i\\
n_1\ n_2\; \cdots \, n_{l-1}\, n_{l} \ n_{l+1}\ \cdots
\end{array}
\qquad
\underset{w_{0m_l,l}^{j\to i}}{\overset{w_{0m_l,l}^{i\to j}}{\rightleftharpoons}}
\qquad
\begin{array}{l}
m_1m_2\cdots m_{l-1} m_{l},\, j\\
n_1\ n_2\; \cdots \, n_{l-1}\, n_{l} \ n_{l+1}\ \cdots
\end{array}
\qquad\qquad
(l\ge 0; i \ne j \in\{1,2,\dots,I\}) \, ,
\label{kin_2}
\ee
where $w_{+m_l,l}^{i\to j}$ is the attachment rate of the monomeric unit $m_l$ to the copolymer chain $m_1m_2\cdots m_{l-1}$ at the location~$l$ of the unit $n_l$ of the template, while the catalyst undergoes its internal transition $i\to j$; $w_{-m_l,l}^{j\to i}$ is the detachment rate of the unit $m_l$ from the copolymer chain upon the reverse transition $j\to i$;  and $w_{0m_l,l}^{i\to j}$ is the rate of the transition $i\to j$ for the catalyst without any monomeric attachment or detachment.  An attachment event increases the chain length by $l \to l+1$, a detachment event decreases it by $l \to l-1$, and the reaction events~(\ref{kin_2}) do not change the length.  The rates depend on the location $l$ along the template sequence, i.e., on the template unit $n_l$.  We may also consider more general dependences of the rates on a larger template subsequence around the location $l$, such as $n_{l-1}n_ln_{l+1}$, or else.  The rates also depend on the concentrations of monomers in the solution surrounding the catalyst and, possibly, on the concentrations of other species activating the process, like the protein EF-G activating the peptide elongation in addition to the monomers (i.e., the aminoacyl-tRNA-EF-Tu molecules) arriving at the ribosome~\cite{GCCR09}.  In particular, the attachment rates $w_{+m,l}^{i\to j}$ are expected to vanish with the concentration $c_m$ of monomers $m$.  The reactions proceed under low conversion conditions, meaning that the rates are low enough to maintain the pools of monomers in the surrounding solution, whereupon the rates can remain invariant during the whole process.  Furthermore, the catalyst is assumed to stay bonded to the template.  

The copolymerization process starts with the following initiation steps:
\be
\begin{array}{l}
i \qquad\qquad\qquad  + \ m_1\\
n_1 \, n_2 \, n_3 \; \cdots 
\end{array}
\qquad
\underset{w_{-m_1,1}^{j \to i}}{\overset{w_{+m_1,1}^{i \to j}}{\rightleftharpoons}}
\qquad
\begin{array}{l}
m_1,\, j\\
n_1 \, n_2 \, n_3 \; \cdots
\end{array}
\qquad
(l = 1; i,j\in\{1,2,\dots,I\}) \, ,
\label{kin0_1}
\ee
and
\be
\quad
\begin{array}{l}
i\\
n_1 \, n_2 \, n_3 \; \cdots
\end{array}
\qquad
\underset{w_{0\emptyset,0}^{j\to i}}{\overset{w_{0\emptyset,0}^{i\to j}}{\rightleftharpoons}}
\qquad
\begin{array}{l}
j\\
n_1 \, n_2 \, n_3 \; \cdots
\end{array}
\qquad\qquad\qquad
(l = 0; i \ne j \in\{1,2,\dots,I\}) \, ,
\label{kin0_2}
\ee
defined with the rates $w_{\pm m_1,1}^{i \to j}$ and $w_{0\emptyset,0}^{i\to j}$.
We also assume that the copolymerization process has no termination steps.

If the attachment rates are large enough with respect to the detachment rates, the copolymer chain grows without bound and its length increases on average in time, so that the catalyst has an infinite processivity along the semi-infinite template.  Otherwise, the copolymer chain remains finite on average and there is no elongation.

We note that simpler processes have been considered in Ref.~\cite{G20}, where $w_{\pm m,l}^{i\to j}=0$ for $i \ne j$ and $w_{0m,l}^{i\to j}$ is independent of the unit $m$.  Here, we consider the more general processes without these simplifications.

\subsection{The kinetic equations}

For a single catalyst bonded to the template and the copy, the process is stochastic with random events of attachments, detachments, and internal transitions of the catalyst.  These random events occur with the reaction rates given in Eqs.~(\ref{kin_1})-(\ref{kin_2}).  The stochastic process is assumed to be a so-called Markov jump process \cite{vK81,Gardiner,S16}.  For such a stochastic process, the system may be found with some probability in either one or another of the following system states:
\be
(\emptyset,0,i) \, ; \quad (m_1,1,i) \, ; \quad (m_1m_2,2,i) \, ; \quad (m_1m_2m_3,3,i) \, ; \quad \dots \quad ; \quad (m_1 m_2 \cdots m_{l-1}m_l,l,i) \, ; \quad \dots \quad
\label{syst-states}
\ee
with $m_l\in\{1,2,\dots,M\}$, $l\in\{0,1,2,3,\dots\}$, $i\in\{1,2,\dots,I\}$, $\emptyset$ denoting the state without any copy yet.  We note that the system state
\be
(m_1 m_2 \cdots m_{l-1}m_l,l,i)  
\qquad \mbox{stands for}\qquad
\begin{array}{l}
m_1m_2\cdots m_{l-1} m_{l},\, i\\
n_1\ n_2\; \cdots \, n_{l-1}\, n_{l} \ n_{l+1}\ \cdots
\end{array}
\ee
in Eqs.~(\ref{kin_1}) and~(\ref{kin_2}).  In this system state, the copy chain has the length $l$ and the catalyst is in its internal state $i$.  The probability to find the system in this state at time $t$ is denoted
\be
P_t(m_1 \cdots m_l,l,i)
\equiv P_t\bigg({m_1 \cdots m_l,\, i \quad\quad\ \atop n_1\, \cdots \, n_l \, n_{l+1}\cdots}\bigg)\, .
\ee 
The time evolution of these probabilities is ruled by the following infinite set of coupled linear ordinary differential equations:
\bea
\frac{\dd}{\dd t}\, P_t(m_1 \cdots m_l ,l,i) &=& \sum_j w_{+m_l,l}^{j\to i} \, P_t(m_1 \cdots m_{l-1} ,l-1,j)
\nonumber\\
&&+\sum_j \sum_{m_{l+1}} w_{-m_{l+1}, l+1}^{j\to i} \, P_t(m_1 \cdots m_l m_{l+1} ,l+1,j)
\nonumber\\
&&+ \sum_{j(\ne i)} w_{0m_l,l}^{j\to i} \, P_t(m_1 \cdots m_l ,l,j)
\nonumber\\
&&- \bigg(\sum_j w_{-m_l,l}^{i\to j} + \sum_j \sum_{m_{l+1}} w_{+m_{l+1},l+1}^{i\to j} + \sum_{j(\ne i)} w_{0m_l,l}^{i\to j}\bigg) P_t(m_1 \cdots m_l ,l,i)
\label{kin_eq}
\eea
for $l\ge 0$, where the positive terms are the gain terms and the negative ones the loss terms.  Moreover, we suppose that $w_{\pm\emptyset,0}^{i\to j}=0$ in the equations for $P_t(\emptyset,0,i)$ with $l=0$.  As a consequence, probability conservation is satisfied, i.e.,
\be
\sum_{l=0}^{\infty} \sum_{i=1}^{I} \sum_{m_1\cdots m_l} P_t(m_1 \cdots m_l ,l,i) = 1
\label{tot-prob}
\ee
for all time $t$.

\subsection{The observable quantities}

For this stochastic process, several observable quantities can be defined to characterize the copolymer chain, its growth, and its sequence.

\subsubsection{Length distribution, mean length, and mean growth velocity}

The probability distribution of the copolymer length at time $t$ is defined as
\be
p_t(l) \equiv \sum_{i=1}^{I} \sum_{m_1\cdots m_l} P_t(m_1 \cdots m_l ,l,i)\, , 
\label{p(l)}
\ee
which obeys the normalization condition $\sum_{l=0}^{\infty} p_t(l)=1$, in consistency with Eq.~(\ref{tot-prob}).
The mean length of the copolymer chain at time $t$ is thus given by
\be
\langle l \rangle_t \equiv \sum_{l=0}^{\infty} l \, p_t(l) \, .
\label{av(l)}
\ee
Accordingly, the mean growth velocity can be defined as
\be
v \equiv \lim_{t\to\infty} t^{-1} \, \langle l \rangle_t \, .
\label{dfn-v}
\ee

\subsubsection{The sequence probabilities}

In the long-time limit, the probability to find the copy with the sequence $m_1\cdots m_l$ given that its length is equal to~$l$ and the catalyst is found in any one of its internal states is defined as
\be
\mu(m_1\cdots m_l;l) \equiv \lim_{t\to\infty} \sum_{i=1}^{I} P_t(m_1 \cdots m_l ,l,i)/p_t(l) \, ,
\label{mu(seq)}
\ee
such that $\sum_{m_1\cdots m_l} \mu(m_1\cdots m_l;l) = 1$.

\subsubsection{The local probabilities of monomeric units in the copy}

As a consequence of Eq.~(\ref{mu(seq)}), the probability to find the monomeric unit $m=m_l$ at the location $l$ in a grown copy sequence is given by
\be
\mu(m;l) \equiv \lim_{L\to\infty} \sum_{m_1\cdots m_{l-1}} \sum_{m_{l+1}\cdots m_L} \mu(m_1\cdots m_{l-1} m \,  m_{l+1}\cdots m_L,L) \, ,
\label{mu(m;l)}
\ee
which is normalized according to $\sum_m \mu(m;l)=1$.

\subsubsection{The mean fractions of monomeric units}

Therefore, the mean fraction of the monomeric unit $m$ anywhere in an arbitrarily long copy sequence can be calculated as
\be
\bar\mu(m) \equiv \lim_{L\to\infty} \frac{1}{L} \sum_{l=1}^{L} \mu(m;l) \, ,
\label{fr(m)}
\ee
satisfying the normalization condition, $\sum_m \bar\mu(m)=1$.

\subsection{The different regimes of the process}

If the template is periodic, we may expect two generic regimes:

\begin{itemize}

\item[(1)] If the attachment rates are too small with respect to the detachment rates for generating the elongation of the copy chain, the mean length converges toward a finite and stationary value, $\lim_{t\to\infty} \langle l \rangle_t = \langle l \rangle_{\rm st}$.  Consistently, the length distribution converges toward a stationary probability distribution, $\lim_{t\to\infty} p_t(l) = p_{\rm st}(l)$, having a mean exponential decay with an exponent $\alpha$: $p_{\rm st}(l)\sim \exp(-\alpha l)$.  This is the stationary regime.

\item[(2)] If the attachment rates are large enough with respect to the detachment rates, there is elongation of the copy chain and its mean length increases linearly in time as $\langle l \rangle_t \simeq vt$ with a positive mean growth velocity $v>0$.  Correspondingly, the length distribution spreads toward arbitrarily large lengths, while becoming broader and broader.  This is the regime of steady growth.

\end{itemize}

If the template is disordered, we may also expect an intermediate regime between (1) and (2), where the mean length manifests a sublinear growth in time like $\langle l \rangle_t \sim t^\gamma$ with an exponent $0 < \gamma < 1$ \cite{D83,BG90}.  This intermediate regime extends between the edge of the stationary regime (1), where the growth is stalled and $\gamma=0$, and the threshold of steady growth, where $\gamma=1$.

\subsection{Matrix formulation}

In order to deal more compactly with the multiple states $i,j=1,2,\dots,I$ of the catalyst, we adopt a matrix formulation.  We introduce the following $I\times I$ matrix of rank one, composed of $I$ identical columns, each containing the probabilities to find the catalyst in the states $i=1,2,\dots,I$:
\be
\boldsymbol{\mathsf P}_t(m_1\cdots m_l,l) \equiv
\left[
\begin{array}{cccc}
P_t(m_1\cdots m_l,l,1) & P_t(m_1\cdots m_l,l,1) & \dots & P_t(m_1\cdots m_l,l,1) \\
P_t(m_1\cdots m_l,l,2) & P_t(m_1\cdots m_l,l,2) & \dots & P_t(m_1\cdots m_l,l,2) \\
\vdots & \vdots & \ddots & \vdots \\
P_t(m_1\cdots m_l,l,I) & P_t(m_1\cdots m_l,l,I) & \dots & P_t(m_1\cdots m_l,l,I)
\end{array}
\right] ,
\label{P-matrix}
\ee
such as its trace,
\be
{\rm tr}\, \boldsymbol{\mathsf P}_t(m_1\cdots m_l,l) = \sum_{i=1}^{I} P_t(m_1\cdots m_l,l,i) \, ,
\label{tr-P-matrix}
\ee
gives the probability to find the copy sequence $m_1\cdots m_l$ of length $l$ and the catalyst in any one of its internal states.

In this formulation, the equations~(\ref{kin_eq}) for $i=1,2,\dots,I$ read
\bea
\frac{\dd}{\dd t}\, \boldsymbol{\mathsf P}_t(m_1\cdots m_l,l) &=& \boldsymbol{\mathsf W}_{+m_l,l}^{\rm c} \cdot \boldsymbol{\mathsf P}_t(m_1 \cdots m_{l-1} ,l-1)
 + \sum_{m_{l+1}} \boldsymbol{\mathsf W}_{-m_{l+1},l+1}^{\rm c} \cdot  \boldsymbol{\mathsf P}_t(m_1 \cdots m_l m_{l+1} ,l+1)
\nonumber\\
&&+ \, \bigg(\boldsymbol{\mathsf W}_{m_l,l}^{0} - \boldsymbol{\mathsf W}_{-m_l,l}^{\rm d} - \sum_{m_{l+1}} \boldsymbol{\mathsf W}_{+m_{l+1},l+1}^{\rm d} \bigg) \cdot\boldsymbol{\mathsf P}_t(m_1 \cdots m_l ,l)
\label{matrix_kin_eq}
\eea
in terms of the following $I\times I$ matrices defined with the rates:
\be
\boldsymbol{\mathsf W}_{\pm m,l}^{\rm c} \equiv
\left[
\begin{array}{cccc}
w_{\pm m,l}^{1\to 1} & w_{\pm m,l}^{2\to 1} & \dots & w_{\pm m,l}^{I\to 1} \\
w_{\pm m,l}^{1\to 2} & w_{\pm m,l}^{2\to 2} & \dots & w_{\pm m,l}^{I\to 2} \\
\vdots & \vdots & \ddots & \vdots \\
w_{\pm m,l}^{1\to I} & w_{\pm m,l}^{2\to I} & \dots & w_{\pm m,l}^{I\to I} \\
\end{array}
\right] ,
\label{Wc-matrix}
\ee
\be
\boldsymbol{\mathsf W}_{\pm m,l}^{\rm d} \equiv
\left[
\begin{array}{cccc}
\sum_{i=1}^{I} w_{\pm m,l}^{1\to i} & 0 & \dots & 0 \\
0 & \sum_{i=1}^{I} w_{\pm m,l}^{2\to i}& \dots & 0\\
\vdots & \vdots & \ddots & \vdots \\
0& 0 & \dots & \sum_{i=1}^{I} w_{\pm m,l}^{I\to i} \\
\end{array}
\right] ,
\label{Wd-matrix}
\ee
and
\be
\boldsymbol{\mathsf W}_{m,l}^{0} \equiv
\left[
\begin{array}{cccc}
- \sum_{i(\ne 1)} w_{0m,l}^{1\to i} & w_{0m,l}^{2\to 1} & \dots & w_{0m,l}^{I\to 1} \\
w_{0m,l}^{1\to 2} & - \sum_{i(\ne 2)} w_{0m,l}^{2\to 2} & \dots & w_{0m,l}^{I\to 2} \\
\vdots & \vdots & \ddots & \vdots \\
w_{0m,l}^{1\to I} & w_{0m,l}^{2\to I} & \dots & - \sum_{i(\ne I)} w_{0m,l}^{I\to I} \\
\end{array}
\right] ,
\label{W0-matrix}
\ee
where $\sum_{i(\ne j)}$ denotes the sum from $i=1$ to $i=I$, omitting the term with $i=j$.

\section{The solutions of the kinetic equations}
\label{sec:sol}

\subsection{The methods used to solve the kinetic equations}

In order to solve the kinetic equations~(\ref{kin_eq}), we use their linearity, so that their general solutions can be expressed as linear superpositions of particular solutions. 

Moreover, we also use the assumptions that the attachment and detachment rates of the reactions~(\ref{kin_1}) only depend on the monomeric unit that is attached or detached, but not on previously incorporated units and that the internal transition rates of the reactions~(\ref{kin_2}) only depend on the ultimate monomeric unit of the copy chain.  These characteristic features of the kinetic process that is here studied, lead to the factorization of the matrices~(\ref{P-matrix}) containing the sequence probabilities into products of matrices associated with each unit of the sequence.  This matrix factorization was pioneered for irreversible multistate free copolymerization processes by Coleman and Fox~\cite{CF63JCP,CF63JACS,CF63JPS} and later shown to extend to reversible processes \cite{G19JCP,G20}.  To give evidence for this factorization, the matrix form~(\ref{matrix_kin_eq}) of the kinetic equations are transformed into a hierarchy of nested equations for
\bea
\boldsymbol{\mathsf P}_t(l) &\equiv& \sum_{m_1\cdots m_l} \boldsymbol{\mathsf P}_t(m_1\cdots m_l,l) \, , \label{P(l)-dfn}\\
\boldsymbol{\mathsf P}_t(m_l,l) &\equiv& \sum_{m_1\cdots m_{l-1}} \boldsymbol{\mathsf P}_t(m_1\cdots m_{l-1}m_l,l) \, , \label{P(m,l)-dfn}\\
\boldsymbol{\mathsf P}_t(m_{l-1}m_l,l) &\equiv& \sum_{m_1\cdots m_{l-2}} \boldsymbol{\mathsf P}_t(m_1\cdots m_{l-2}m_{l-1}m_l,l) \, , \label{P(mm',l)-dfn}\\
&\vdots& \nonumber
\eea
giving the probabilities of the subsequences $\{m_l,m_{l-1}m_l,m_{l-2}m_{l-1}m_l,\dots\}$ at the growing tip of the copy.  We note that the matrices~(\ref{P(l)-dfn}), (\ref{P(m,l)-dfn}), (\ref{P(mm',l)-dfn}), $\dots$ satisfy the following successive identities
\bea
\boldsymbol{\mathsf P}_t(l) &=& \sum_{m_l} \boldsymbol{\mathsf P}_t(m_l,l) \, , \label{P(l)-P(m,l)}\\
\boldsymbol{\mathsf P}_t(m_l,l) &=& \sum_{m_{l-1}} \boldsymbol{\mathsf P}_t(m_{l-1}m_l,l) \, , \label{P(m,l)-P(mm',l)}\\
\boldsymbol{\mathsf P}_t(m_{l-1}m_l,l) &=& \sum_{m_{l-2}} \boldsymbol{\mathsf P}_t(m_{l-2}m_{l-1}m_l,l) \, , \label{P(mm',l)-P(mm'm'',l)}\\
&\vdots& \nonumber
\eea
so that the equations they obey are nested in the sense that the equation for any subsequence $m_{l-r+1}\cdots m_{l-1}m_l$ of length $r$ can be deduced from the equations for longer subsequences $m_{l-s+1}\cdots m_{l-1}m_l$ with $s>r$ by summing over $m_{l-s+1}\cdots m_{l-r}$.

Furthermore, we have that the length distribution is given by the trace
\be
p_t(l) = {\rm tr} \, \boldsymbol{\mathsf P}_t(l)
\label{p(l)-trace}
\ee
of the first matrix~(\ref{P(l)-dfn}) of the hierarchy.  Solving their equations should thus provide the determination of the mean growth velocity.

If the property of factorization holds, the equations for the matrices~(\ref{P(l)-dfn}), (\ref{P(m,l)-dfn}), (\ref{P(mm',l)-dfn}), $\dots$ should all be solved together, which is the case as shown here below.

\subsection{Decomposition into particular solutions}

We consider particular solutions of the kinetic matrix equations~(\ref{matrix_kin_eq}) like Fourier modes of wave number $q$ with an exponential dependence on time:
\be
\boldsymbol{\mathsf P}_t({\bf m},l) \sim {\rm e}^{s_qt+\imath ql} \, \boldsymbol{\mathsf G}_q({\bf m},l)
\label{q-modes}
\ee
with ${\bf m}=m_{l-r+1}\cdots m_{l-1}m_l$ ($r=1,2,3,\dots$), $-\pi < q \le +\pi$, and $\imath\equiv\sqrt{-1}$.  We note that the quantities $\boldsymbol{\mathsf G}_q({\bf m},l)$ still depend on the location $l$, because the template is in general a disordered medium.

In the regime of steady growth, the exponential rate $s_q$ should depend on the wave number according to
\be
s_q = - \imath \, v \, q + o(q) \, , 
\label{s_q}
\ee
where $v$ is the mean growth velocity~(\ref{dfn-v}) and $o(q)$ is a function of $q$ such that $\lim_{q\to 0} o(q)/q = 0$. Hence, we should have $\exp(s_qt+\imath ql)=\exp[\imath q(l-vt)+o(q)]$, which indeed represents a solution moving along the template with the mean velocity $v$.

During elongation, the length distribution $p_t(l)$ broadens and spreads over longer and longer length scales.  Therefore, in the long-time limit $t\to\infty$, the particular solutions~(\ref{q-modes}) that contribute the most should have their wave number decreasing to zero, $q\to 0$.  In this regard, they can be expanded as
\be
\boldsymbol{\mathsf G}_q({\bf m},l) = \boldsymbol{\Psi}_l({\bf m}) + q \, \boldsymbol{\Psi}_l^{\prime}({\bf m}) + o(q)
\label{G_q}
\ee
with
\be
\boldsymbol{\Psi}_l({\bf m}) \equiv \lim_{q\to 0} \boldsymbol{\mathsf G}_q({\bf m},l) \, , \label{Psi-dfn}
\ee
and
\be
\boldsymbol{\Psi}_l^{\prime}({\bf m}) \equiv \lim_{q\to 0} \frac{\dd}{\dd q} \, \boldsymbol{\mathsf G}_q({\bf m},l) \, . \label{Psi'-dfn}
\ee

Since Eqs.~(\ref{P(l)-P(m,l)}), (\ref{P(m,l)-P(mm',l)}), (\ref{P(mm',l)-P(mm'm'',l)}), $\dots$ hold, we have that
\bea
\boldsymbol{\Psi}_l &=& \sum_{m_l} \boldsymbol{\Psi}_l(m_l) \, , \label{Psi(l)-Psi(m,l)}\\
\boldsymbol{\Psi}_l(m_l) &=& \sum_{m_{l-1}} \boldsymbol{\Psi}_l(m_{l-1}m_l) \, , \label{Psi(m,l)-Psi(mm',l)}\\
\boldsymbol{\Psi}_l(m_{l-1}m_l) &=& \sum_{m_{l-2}} \boldsymbol{\Psi}_l(m_{l-2}m_{l-1}m_l) \, , \label{Psi(mm',l)-Psi(mm'm'',l)}\\
&\vdots& \nonumber
\eea
Similar relations hold for $\boldsymbol{\Psi}_l^{\prime}$, $\boldsymbol{\Psi}_l^{\prime}(m_l)$, $\dots$

The equations for $\boldsymbol{\mathsf G}_q(l)$, $\boldsymbol{\mathsf G}_q(m_l,l)$, $\dots$ and $\boldsymbol{\Psi}_l$, $\boldsymbol{\Psi}_l(m_l)$, $\dots$ are given and solved in Appendix~\ref{AppA}.

\subsection{Matrix factorization}

Since the attachment and detachment rates do not depend on previously incorporated monomeric units and the internal transition rates only depend on the ultimate monomeric unit of the copolymer chain, the following ansatz of matrix factorization is satisfied for the studied kinetic process, as proved in Appendix~\ref{AppA}:
\be
\boldsymbol{\Psi}_l(m_{l-r+1}\cdots m_{l-1}m_l)=\boldsymbol{\mathsf Y}_{m_l,l} \cdot \boldsymbol{\mathsf Y}_{m_{l-1},l-1} \cdots \boldsymbol{\mathsf Y}_{m_{l-r+1},l-r+1} \cdot \boldsymbol{\Psi}_{l-r}
\qquad (r=1,2,3,\dots)
\label{factorization}
\ee
in terms of some $I\times I$ matrices $\boldsymbol{\mathsf Y}_{m,l}$, depending on the monomeric unit $m$ that is found at the location $l$ along the copy sequence.  The factorization ansatz~(\ref{factorization}) is known to hold for simpler copolymerization processes than those of this study \cite{CF63JCP,CF63JACS,CF63JPS,G19JCP,G20}.
 
\subsection{The solution in the growth regime}

\subsubsection{The backward and forward iterations}

The factorization ansatz~(\ref{factorization}) can be substituted into the equations for $\boldsymbol{\Psi}_l$, $\boldsymbol{\Psi}_l(m_l)$, $\dots$.  As shown in Appendix~\ref{AppA}, we obtain the results:
\be
\boldsymbol{\mathsf Y}_{m_l,l} = \left( \boldsymbol{\mathsf V}_l - \boldsymbol{\mathsf W}^{0}_{m_l,l} + \boldsymbol{\mathsf W}^{\rm d}_{-m_l,l}\right)^{-1} \cdot \boldsymbol{\mathsf W}^{\rm c}_{+m_l,l}
\label{Y_ml}
\ee
in terms of the $I\times I$ matrices $\boldsymbol{\mathsf V}_l$, obeying the following {\it backward iteration}:
\be
\boxed{
\boldsymbol{\mathsf V}_{l-1}= \sum_{m_l} \boldsymbol{\mathsf W}^{\rm d}_{+m_l,l} - \sum_{m_l} \boldsymbol{\mathsf W}^{\rm c}_{-m_l,l} \cdot \left( \boldsymbol{\mathsf V}_l - \boldsymbol{\mathsf W}^{0}_{m_l,l} + \boldsymbol{\mathsf W}^{\rm d}_{-m_l,l}\right)^{-1} \cdot \boldsymbol{\mathsf W}^{\rm c}_{+m_l,l}
}
\label{backward_iter}
\ee
along the template sequence.

Once the matrices $\boldsymbol{\mathsf V}_l$ are known, the matrices $\boldsymbol{\mathsf Y}_{m_l,l}$ are also, because of Eq.~(\ref{Y_ml}), so that the quantities~(\ref{factorization}) are obtained.  In particular, if we take Eq.~(\ref{factorization}) with $r=1$, we have that
\be
\boldsymbol{\Psi}_l(m_l)=\boldsymbol{\mathsf Y}_{m_l,l} \cdot \boldsymbol{\Psi}_{l-1} \, .
\label{Psi_ml-Y_ml}
\ee
Inserting the latter relation into Eq.~(\ref{Psi(l)-Psi(m,l)}), we find the {\it forward iteration}:
\be
\boxed{
\boldsymbol{\Psi}_l=\boldsymbol{\mathsf R}_l \cdot \boldsymbol{\Psi}_{l-1}
\qquad\mbox{with}\qquad
\boldsymbol{\mathsf R}_l \equiv \sum_{m_l} \boldsymbol{\mathsf Y}_{m_l,l}
}
\label{forward_iter}
\ee
along the template.  The forward iteration determines the following $I\times I$ matrices of rank one:
\be
\boldsymbol{\Psi}_l =
\left[
\begin{array}{cccc}
\psi_l^1 & \psi_l^1 & \dots & \psi_l^1 \\
\psi_l^2 & \psi_l^2 & \dots & \psi_l^2 \\
\vdots & \vdots & \ddots & \vdots \\
\psi_l^I & \psi_l^I & \dots & \psi_l^I
\end{array}
\right] .
\label{Psi-matrix}
\ee

The matrices $\boldsymbol{\mathsf V}_l$ and $\boldsymbol{\Psi}_l$ obey the remarkable invariance property:
\be
{\rm tr}(\boldsymbol{\mathsf V}_{l-1} \cdot \boldsymbol{\Psi}_{l-1}) = {\rm tr}(\boldsymbol{\mathsf V}_l \cdot \boldsymbol{\Psi}_l) \equiv C
\qquad \mbox{for}\qquad l=1,2,3,\dots \, ,
\label{Constant}
\ee
so that the quantity $C$ is constant along the template.

\subsubsection{The mean growth velocity}

Having computed the matrices $\boldsymbol{\mathsf V}_l$ and $\boldsymbol{\Psi}_l$ with the backward and forward iterations (\ref{backward_iter}) and (\ref{forward_iter}), respectively, the mean growth velocity~(\ref{dfn-v}) is given by
\be
v = \frac{1}{\langle \tau_l \rangle} \, ,
\qquad\mbox{where}\qquad
\tau_l = \frac{1}{C} \, {\rm tr} \, \boldsymbol{\Psi}_l
\label{v-formula}
\ee
is the mean local dwell time of the catalyst at the location $l$ of the template and $\langle\tau_l \rangle \equiv \lim_{L\to\infty} L^{-1} \sum_{l=1}^L \tau_l$ is its statistical average, $C$~being the constant defined in Eq.~(\ref{Constant}).  

The mean growth velocity is positive $v>0$ if the growth is linear in time as $\langle l\rangle_t\simeq vt$, but equal to zero $v=0$ if the growth is sublinear in time like $\langle l\rangle_t\sim t^\gamma$ or if the growth is stalled.

\subsubsection{The sequence probabilities}

Using the factorization~(\ref{factorization}) and Eq.~(\ref{tr-P-matrix}), the sequence probability~(\ref{mu(seq)}) is given in the long-time limit by
\be
\mu(m_{l-r+1}\cdots m_{l-1}m_l;l) = \frac{1}{{\rm tr}\, \boldsymbol{\Psi}_l} \, {\rm tr}(\boldsymbol{\mathsf Y}_{m_l,l} \cdot \boldsymbol{\mathsf Y}_{m_{l-1},l-1} \cdots \boldsymbol{\mathsf Y}_{m_{l-r+1},l-r+1} \cdot \boldsymbol{\Psi}_{l-r})
\qquad\mbox{for}\qquad 
r=1,2,3,\dots
\label{mu(seq)-formula}
\ee
The normalization condition $\sum_{m_{l-r+1}\cdots m_{l-1}m_l} \mu(m_{l-r+1}\cdots m_{l-1}m_l;l) =1$ is satisfied, because of the forward iteration~(\ref{forward_iter}).

\subsubsection{The local probabilities of monomeric units}

The probability~(\ref{mu(m;l)}) of finding the monomeric unit $m=m_l$ at the location $l$ along a template of length $L$ can thus be computed with
\be
\mu(m;l) = \lim_{L\to\infty} \frac{1}{{\rm tr}\, \boldsymbol{\Psi}_L} \, {\rm tr}(\boldsymbol{\mathsf S}_l \cdot \boldsymbol{\mathsf Y}_{m,l}  \cdot \boldsymbol{\Psi}_{l-1}) \, ,
\label{mu(m;l)-formula}
\ee
where the matrices
\be
\boldsymbol{\mathsf S}_l \equiv \boldsymbol{\mathsf R}_L \cdots \boldsymbol{\mathsf R}_{l+1}
\label{S-matrix}
\ee
are obtained by chain multiplication of the matrices $\boldsymbol{\mathsf R}_l$ defined in Eq.~(\ref{forward_iter}) and using the fact that $\boldsymbol{\Psi}_{l-1} = \boldsymbol{\mathsf R}_{l-1}\cdots \boldsymbol{\mathsf R}_1\cdot\boldsymbol{\Psi}_0$.
Subsequently, the mean fractions of monomeric units $m\in\{1,2,\dots,M\}$ in the copy are given by Eq.~(\ref{fr(m)}).

We note that the matrices~(\ref{S-matrix}) satisfy the recurrence $\boldsymbol{\mathsf S}_{l-1}=\boldsymbol{\mathsf S}_l \cdot \boldsymbol{\mathsf R}_l$, starting from $\boldsymbol{\mathsf S}_L=\boldsymbol{\mathsf 1}$ (where $\boldsymbol{\mathsf 1}$ is the $I\times I$ identity matrix) and ending with $\boldsymbol{\mathsf S}_0 = \boldsymbol{\mathsf R}_L \cdot \boldsymbol{\mathsf R}_{L-1} \cdots \boldsymbol{\mathsf R}_1$.  Moreover, we have the property that $\boldsymbol{\mathsf S}_{l-1}\cdot\boldsymbol{\Psi}_{l-1}=\boldsymbol{\mathsf S}_l\cdot\boldsymbol{\Psi}_l$, so that $\boldsymbol{\Psi}_L=\boldsymbol{\mathsf S}_0\cdot\boldsymbol{\Psi}_0$.  

If we assume that the template is periodic of period $L$ and if we consider the periodic boundary conditions $\boldsymbol{\Psi}_L=\boldsymbol{\Psi}_0$, we infer that the matrix $\boldsymbol{\mathsf S}_0$ should have an eigenvalue equal to one among its eigenvalues, i.e., the property that $\boldsymbol{\mathsf S}_0\cdot\boldsymbol{\Psi}_0=\boldsymbol{\Psi}_0$.  Under such circumstances, the process is propagative along the template and the copy grows as time increases.

\subsubsection{Relations to the solutions for simpler processes}

In summary, the backward and forward iterations (\ref{backward_iter}) and (\ref{forward_iter}) provide the exact asymptotic solution of the kinetic equations~(\ref{kin_eq}) ruling the stochastic process~(\ref{kin_1})-(\ref{kin_2}).

We note that, in the simpler processes where $w_{\pm m,l}^{i\to j}=0$ for $i\ne j$, the matrices~(\ref{Wc-matrix}) and~(\ref{Wd-matrix}) are equal to each other, so that the iterations (\ref{backward_iter}) and (\ref{forward_iter}) reduce to those of the  template-directed multistate copolymerization processes of Ref.~\cite{G20} with $\boldsymbol{\mathsf W}_{\pm m,l}^{\rm c}=\boldsymbol{\mathsf W}_{\pm m,l}^{\rm d}\equiv \boldsymbol{\mathsf W}_{\pm m,l}$.  If the template is furthermore uniform (i.e., periodic of period $L=1$), the equations reduce to those of the free multistate copolymerization processes studied in Ref.~\cite{G19JCP}.

In the case where there is a single internal state (i.e., $I=1$), the transition matrices~(\ref{W0-matrix}) are equal to zero $\boldsymbol{\mathsf W}_{m,l}^{0}=0$, so that all the matrices reduce to scalars.  In particular, the matrices $\boldsymbol{\mathsf V}_l$ reduce to the scalar quantities $v_l=x_l=1/\tau_l$ and the backward iteration~(\ref{backward_iter}) becomes Eq.~(3) of Ref.~\cite{G16PRL}, while the matrices $\boldsymbol{\Psi}_l$ and $\boldsymbol{\mathsf R}_l$ respectively reduce to the scalar quantities $\psi_l$ and $r_l=\psi_l/\psi_{l-1}=v_{l-1}/v_l$, which play a redundant role with respect to the local velocities $v_l$.

\subsection{The solution in the stationary regime}

If the attachment rates are too small with respect to the detachment rates, the growth is stalled.  In this stationary regime, the backward iteration converges toward matrices $\boldsymbol{\mathsf V}_l$ of rank lower than their size $I$, such that $\det\boldsymbol{\mathsf V}_l=0$ and $C={\rm tr} \left(\boldsymbol{\mathsf V}_l \cdot \boldsymbol{\Psi}_l\right)=0$ for $l=1,2,3, \dots$, because the matrices $\boldsymbol{\Psi}_l$ have the form~(\ref{Psi-matrix}). Consequently, the mean growth velocity is equal to zero, $v=0$, as it should.  In this regime, the length distribution reaches stationarity and, according to Eq.~(\ref{p(l)-trace}), it is given by
\be
p_{\rm st}(l) = \lim_{t\to\infty} {\rm tr} \, \boldsymbol{\mathsf P}_t(l) = \frac{{\rm tr} \, \boldsymbol{\Psi}_l}{\sum_{l=1}^{\infty} {\rm tr} \, \boldsymbol{\Psi}_l} \, ,
\label{p_st(l)}
\ee
since the wave number vanishes ($q\to 0$) in the long-time limit ($t\to \infty$).

Typically, the stationary length distribution decreases exponentially as $p_{\rm st}(l)\sim\exp(-\alpha l)$ with a positive exponent $\alpha>0$.  This exponent goes to zero at the onset of growth, which is approached from the stationary regime when the attachment rates become larger and larger with respect to the detachment rates.  Above the onset of growth, the mean length is no longer stationary and it increases with time.

If the template is disordered, the increase of the mean length may be sublinear in time, growing as $\langle l\rangle_t\sim t^\gamma$ ($0<\gamma<1$), as explained in Appendix~\ref{AppB}.  For larger values of the attachment rates, the increase can be linear in time going as $\langle l\rangle_t\simeq vt$ with a positive mean growth velocity $v>0$.  Under such circumstances, the onset of growth (where the growth is still stalled) corresponds to the value $\gamma=0$ of the exponent $\gamma$ and the threshold of linear growth to $\gamma=1$.  The values of the rates are typically different at the onset of growth ($\gamma=0$) and at the threshold of linear growth ($\gamma=1$) if the template is disordered.  Otherwise, if the template has a finite period, the regime of sublinear growth in time does not exist and the mean growth velocity directly becomes positive beyond the stationary regime where growth is stalled.

\section{Numerical example}
\label{sec:num_example}

In this section, the method based on the backward and forward iterations (\ref{backward_iter}) and (\ref{forward_iter}) to solve the kinetic equations~(\ref{kin_eq}) is compared to the method of numerical simulation with Gillespie's algorithm for the corresponding stochastic process~\cite{G76,G77}.  The application of Gillespie's algorithm to this template-directed multistate copolymerization process defined by the reactions~(\ref{kin_1}) and~(\ref{kin_2}) is summarized in Appendix~\ref{AppC}.

\subsection{The model}

The vehicle of this comparison is the following model with two types of monomeric units for the template and the copy ($M=N=2$) and a catalyst having two internal states ($I=2$).  Furthermore, the dependence of the rates on the location $l$ along the template is expressed in terms of the template unit $n_l$, to which the copy unit $m_l$ is paired, i.e., the rates of the model are given by $w_{\pm m_l,l}^{i\to j}=w_{\pm m_l,n_l}^{i\to j}$ and $w_{0m_l,l}^{i\to j}=w_{0m_l,n_l}^{i\to j}$.  Moreover, the attachment rates are assumed to be proportional to the concentration of the monomer that is attached, while the detachment rates are not, and the transition rates are independent of the monomeric unit in the copy.  The rates are thus taken to be
\be
w_{+m,n}^{i\to j} = k_{+m,n}^{i\to j} \, c_m \, , \qquad
w_{-m,n}^{i\to j} = k_{-m,n}^{i\to j} \, , \qquad
\mbox{and}\qquad
w_{0m,n}^{i\to j} = k_{0,n}^{i\to j}
\label{model-rates}
\ee
for $m,n=1,2$ and $i,j=1,2$ with the following rate constants:
\bea
&& k_{+1,1}^{1\to 1}=3 \, , \quad
      k_{+1,2}^{1\to 1}=2 \, , \quad
      k_{+2,1}^{1\to 1}=5 \, , \quad
      k_{+2,2}^{1\to 1}=4 \, , \quad
      k_{+1,1}^{1\to 2}=1 \, , \quad
      k_{+1,2}^{1\to 2}=2 \, , \quad
      k_{+2,1}^{1\to 2}=2 \, , \quad
      k_{+2,2}^{1\to 2}=5 \, , \quad
\nonumber\\
&&  k_{+1,1}^{2\to 1}=1 \, , \quad
      k_{+1,2}^{2\to 1}=3 \, , \quad
      k_{+2,1}^{2\to 1}=6 \, , \quad
      k_{+2,2}^{2\to 1}=3 \, , \quad
      k_{+1,1}^{2\to 2}=6 \, , \quad
      k_{+1,2}^{2\to 2}=4 \, , \quad
      k_{+2,1}^{2\to 2}=4 \, , \quad
      k_{+2,2}^{2\to 2}=2 \, , \quad
\label{attach_rate_csts}\\
&&  k_{-1,1}^{1\to 1}=2 \, , \quad
      k_{-1,2}^{1\to 1}=1 \, , \quad
      k_{-2,1}^{1\to 1}=2 \, , \quad
      k_{-2,2}^{1\to 1}=6 \, , \quad
      k_{-1,1}^{1\to 2}=4 \, , \quad
      k_{-1,2}^{1\to 2}=1 \, , \quad
      k_{-2,1}^{1\to 2}=3 \, , \quad
      k_{-2,2}^{1\to 2}=1 \, , \quad
\nonumber\\
&&  k_{-1,1}^{2\to 1}=5 \, , \quad
      k_{-1,2}^{2\to 1}=2 \, , \quad
      k_{-2,1}^{2\to 1}=1 \, , \quad
      k_{-2,2}^{2\to 1}=1 \, , \quad
      k_{-1,1}^{2\to 2}=1 \, , \quad
      k_{-1,2}^{2\to 2}=2 \, , \quad
      k_{-2,1}^{2\to 2}=1 \, , \quad
      k_{-2,2}^{2\to 2}=3 \, , \quad
\label{detach_rate_csts}\\
&&  k_{0,1}^{1\to 2}=4 \, , \quad
      k_{0,2}^{1\to 2}=1 \, , \quad
        k_{0,1}^{2\to 1}=2 \, , \quad
      k_{0,2}^{2\to 1}=3 \, .
\label{trans_rate_csts}
\eea
These rates determine the matrices $\boldsymbol{\mathsf W}_{\pm m,n}^{\rm c}$, $\boldsymbol{\mathsf W}_{\pm m,n}^{\rm d}$, and~$\boldsymbol{\mathsf W}_{n}^{0}$ according to Eqs.~(\ref{Wc-matrix}), (\ref{Wd-matrix}), and~(\ref{W0-matrix}), respectively, where the subscript~$l$ is replaced with $n=n_l$ and $I=2$.

\subsection{The backward and forward iterations}

Here, the backward iteration~(\ref{backward_iter}) is given by the {\it iterated matrix function system} (IMFS):
\be
\boldsymbol{\mathsf V}_{l-1}= \boldsymbol{\mathsf F}_{n_l} (\boldsymbol{\mathsf V}_l)
\label{IMFS}
\ee
in terms of the following system of two matrix functions,
\be
\boldsymbol{\mathsf F}_{n} (\boldsymbol{\mathsf V}) \equiv \sum_{m=1}^{2} \boldsymbol{\mathsf W}^{\rm d}_{+m,n} - \sum_{m=1}^{2} \boldsymbol{\mathsf W}^{\rm c}_{-m,n} \cdot \left( \boldsymbol{\mathsf V} - \boldsymbol{\mathsf W}^{0}_{n} + \boldsymbol{\mathsf W}^{\rm d}_{-m,n}\right)^{-1} \cdot \boldsymbol{\mathsf W}^{\rm c}_{+m,n}
\qquad\mbox{for} \quad n=1,2 \, .
\label{MF}
\ee
This iteration runs backward along the template sequence $n_1n_2\cdots n_{l-1}n_l\cdots$, which is taken as a Bernoulli sequence with the probabilities $\nu(n)=0.5$ for $n=1,2$.  Therefore, the $2\times 2$ matrix~$\boldsymbol{\mathsf V}_{l-1}$ is determined from the $2\times 2$ matrix~$\boldsymbol{\mathsf V}_l$, depending on the template unit $n_l\in\{1,2\}$ at the location $l$.

If the template sequence is periodic with $n_{l+L}=n_l$ for $l=1,2,\dots,L$ and, thus, forms a loop, numerical convergence can be achieved by running the backward iteration~(\ref{IMFS}) several times around the loop.  The number of passages around the loop is denoted $N_{\rm loop}$.  We note that this iteration converges exponentially fast in both the stationary and the growth regimes.  The rate of exponential convergence of the iteration~(\ref{IMFS}) vanishes at the edge of the stationary regime, where the growth of the copolymer chain is stalled.  Otherwise, it is positive.

After numerical convergence toward the sequence of $2\times 2$ matrices $\{\boldsymbol{\mathsf V}_l\}_{l=1}^L$, the matrices~(\ref{Y_ml}) can be computed, which here read as follows,
\be
\boldsymbol{\mathsf Y}_{m,l} = \left( \boldsymbol{\mathsf V}_l - \boldsymbol{\mathsf W}^{0}_{n_l} + \boldsymbol{\mathsf W}^{\rm d}_{-m,n_l}\right)^{-1} \cdot \boldsymbol{\mathsf W}^{\rm c}_{+m,n_l} \, ,
\label{Y_ml-Ex}
\ee
giving the matrices $\boldsymbol{\mathsf R}_l \equiv \sum_{m=1}^{2} \boldsymbol{\mathsf Y}_{m,l}$ of the forward iteration~(\ref{forward_iter}).  The latter is run forward along the template sequence, starting from some initial condition $\boldsymbol{\Psi}_0$ and generating the sequence of $2\times 2$ matrices $\{\boldsymbol{\Psi}_l\}_{l=1}^L$.  The matrices $\boldsymbol{\Psi}_l$ have the form~(\ref{Psi-matrix}) with $I=2$ and they are thus of rank one.

\subsection{The stationary regime}

This regime exists if the monomer concentrations $c_1$ and $c_2$ are low enough for the attachment events to have a smaller rate than the detachment events.  In this regime, the backward iteration~(\ref{IMFS})-(\ref{MF}) converges toward matrices of rank one such that
\be
\boldsymbol{\mathsf V}_l = \left[
\begin{array}{rr}
v_l^1 & -v_l^2 \\
-v_l^1 & v_l^2
\end{array}
\right]
\qquad\mbox{with}\qquad
v_l^1 = V_l^{11} = - V_l^{21}
\qquad\mbox{and}\qquad
v_l^2 = V_l^{22} = - V_l^{12} \, .
\label{V_l-degen}
\ee
Consequently, the mean growth velocity~(\ref{v-formula}) is equal to zero because the matrices $\boldsymbol{\Psi}_l$ are also of rank one, but of the following form
\be
\boldsymbol{\Psi}_l =
\left[
\begin{array}{cc}
\psi_l^1 & \psi_l^1 \\
\psi_l^2 & \psi_l^2 
\end{array}
\right] ,
\label{Psi-matrix-2}
\ee
so that $C={\rm tr}(\boldsymbol{\mathsf V}_l \cdot \boldsymbol{\Psi}_l)={\rm tr}(\boldsymbol{\Psi}_l\cdot\boldsymbol{\mathsf V}_l)=0$.  Therefore, the growth of the copy is stalled.  

\begin{figure}[h]
\includegraphics[width=8.5cm]{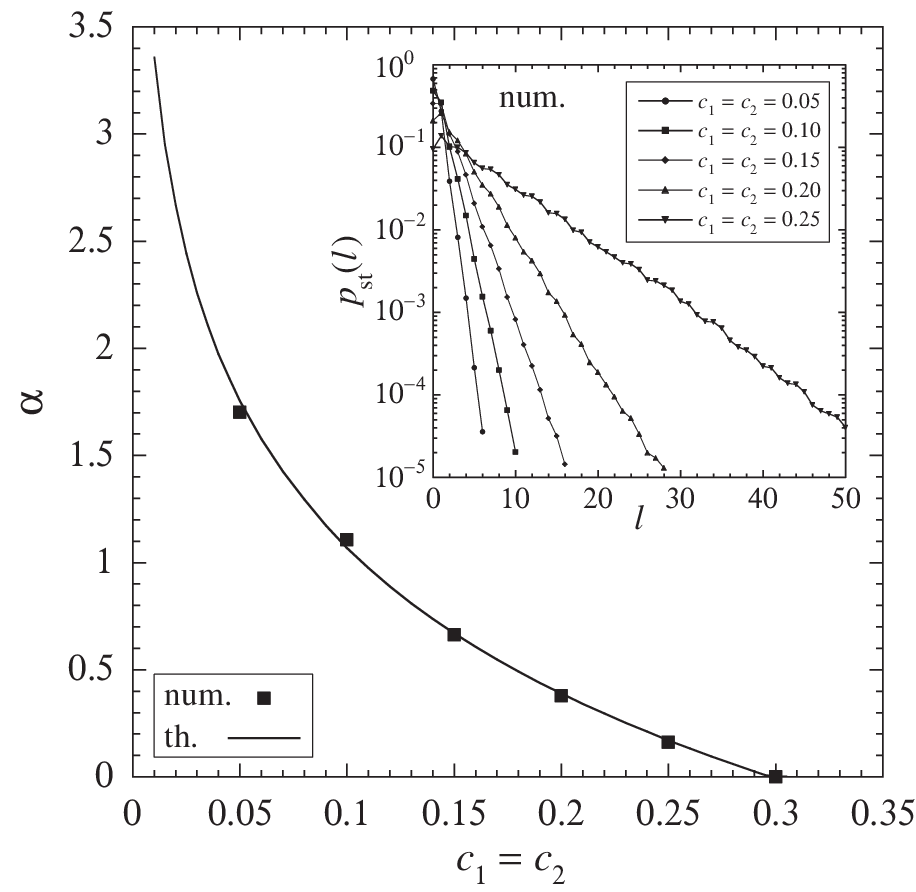}
\caption{Exponent $\alpha$ of the exponential decrease of the stationary length distribution $p_{\rm st}(l)$ versus equal monomer concentrations $c_1=c_2$. The dots depict the numerical results of Gillespie's algorithm and the solid line the theoretical predictions from Eq.~(\ref{p_st(l)}) and ${\rm tr}\, \boldsymbol{\Psi}_l =\psi_l^1+\psi_l^2\sim\exp(-\alpha l)$.  The dots are obtained from the length distributions $p_{\rm st}(l)$ shown in the inset.  These distributions are computed in the stationary regime by simulating a long random trajectory $l(t)$ with $N_{\rm step}=10^7$ steps of Gillespie's algorithm and counting how often the length is found at some value $l$ during the simulation of the trajectory.  The template is a Bernoulli sequence of probabilities $\nu(1)=\nu(2)=0.5$.}
\label{fig1}
\end{figure}

As a corollary, there is an exponential decrease such that ${\rm tr}\,\boldsymbol{\Psi}_l \sim\exp(-\alpha l)$ with $\alpha >0$, giving the stationary length distribution~(\ref{p_st(l)}).  Figure~\ref{fig1} shows the exponent $\alpha$ of that distribution as a function of equal concentrations $c_1=c_2$ for the two monomers.  In Fig.~\ref{fig1}, the solid line depicts the theoretical predictions of the iterative method and the dots the results of the numerical simulation with Gillespie's algorithm.  The inset of Fig.~\ref{fig1} gives several examples of stationary length distributions obtained with Gillespie's algorithm, showing their exponential decrease as $p_{\rm st}(l)\sim \exp(-\alpha l)$ on average.  We observe the good agreement between the theoretical and numerical results.  The exponent $\alpha$ decreases as the equal concentrations $c_1=c_2$ increases, to vanish at the edge of the stationary regime, which thus extends over the following concentrations:
\be
\mbox{the stationary regime:} \qquad  c_1=c_2 < 0.2973 \pm 0.0002 \, .
\label{c1=c2_stat_reg}
\ee

\subsection{The growth regime}

In this regime, the backward iteration~(\ref{IMFS})-(\ref{MF}) converges toward matrices  $\{\boldsymbol{\mathsf V}_l\}_{l=1}^L$ of rank two.  The matrix $\boldsymbol{\mathsf S}_0$ given by Eq.~(\ref{S-matrix}) with $l=0$ has an eigenvalue equal to one, so that we have $\boldsymbol{\Psi}_L=\boldsymbol{\Psi}_0=\boldsymbol{\mathsf S}_0\cdot\boldsymbol{\Psi}_0$ with
\be
\frac{\psi_0^1}{\psi_0^2} = \frac{S_0^{12}}{1-S_0^{11}} = \frac{1-S_0^{22}}{S_0^{21}}
\qquad\mbox{and}\qquad
\psi_0^1+\psi_0^2 = 1 \, .
\label{psi0-S2}
\ee
Consequently, the growth of the copy becomes possible and the grown copy can be characterized by the statistics of the monomeric units composing its sequence.

\begin{figure}[h]
\includegraphics[width=9cm]{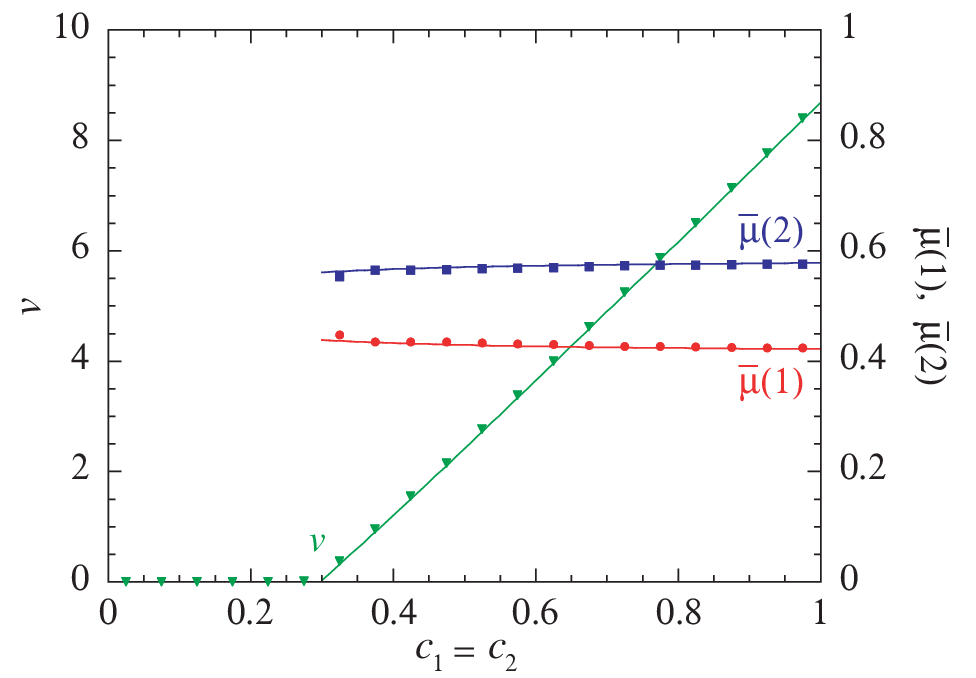}
\caption{The mean growth velocity and the mean fractions $\bar\mu(1)$ and $\bar\mu(2)$ of monomeric units in the grown copy versus equal monomer concentrations $c_1=c_2$, showing how these quantities behave in the transition between the stationary and growth regimes. The template is a Bernoulli sequence of probabilities $\nu(1)=\nu(2)=0.5$ and length $L=10^4$.  The symbols are the numerical results of Gillespie's algorithm with $N_{\rm stat}=10^4$ and the solid lines the theoretical predictions of the iterative method with $N_{\rm loop}=10^2$.}
\label{fig2}
\end{figure}

Figure~\ref{fig2} shows the mean growth velocity~(\ref{dfn-v}) and the mean fractions~(\ref{fr(m)}) of monomeric units $m=1,2$ in the copy as functions of equal concentrations $c_1=c_2$ for the two types of monomers in the solution surrounding the catalyst.  We observe therein the onset of steady growth where $v>0$, which exists for the following concentrations:
\be
\mbox{the steady-growth regime:} \qquad  c_1=c_2 > 0.2976 \pm 0.0002 \, .
\label{c1=c2_st_growth_reg}
\ee
In Fig.~\ref{fig2}, the dots are obtained with Gillespie's algorithm and the solid lines are given by the iterative method, using Eq.~(\ref{v-formula}) to compute the mean growth velocity and Eq.~(\ref{fr(m)}) with Eq.~(\ref{mu(m;l)-formula}) for the fractions of the two types of monomeric units in the copy.  We see in Fig.~\ref{fig2} that the mean growth velocity increases with the monomer concentrations, since the attachment rates are proportional to the concentrations according to Eq.~(\ref{model-rates}), so that they increase with respect to the detachment rates, which stay constant.  Here also, there is good agreement between the results of the iterative method and Gillespie's algorithm.

We note that, in the tiny interval $0.2973 \pm 0.0002 \leq c_1=c_2 \leq 0.2976 \pm 0.0002$ in between the stationary and the steady-growth regimes (\ref{c1=c2_stat_reg}) and~(\ref{c1=c2_st_growth_reg}), there is a regime a sublinear growth in time with $\langle l \rangle_t\sim t^\gamma$, as confirmed with the calculation of the exponent $\gamma$ in Appendix~\ref{AppB}, showing that $0\leq \gamma \leq 1$ in this interval of concentrations.

\begin{figure}[h]
\includegraphics[width=9cm]{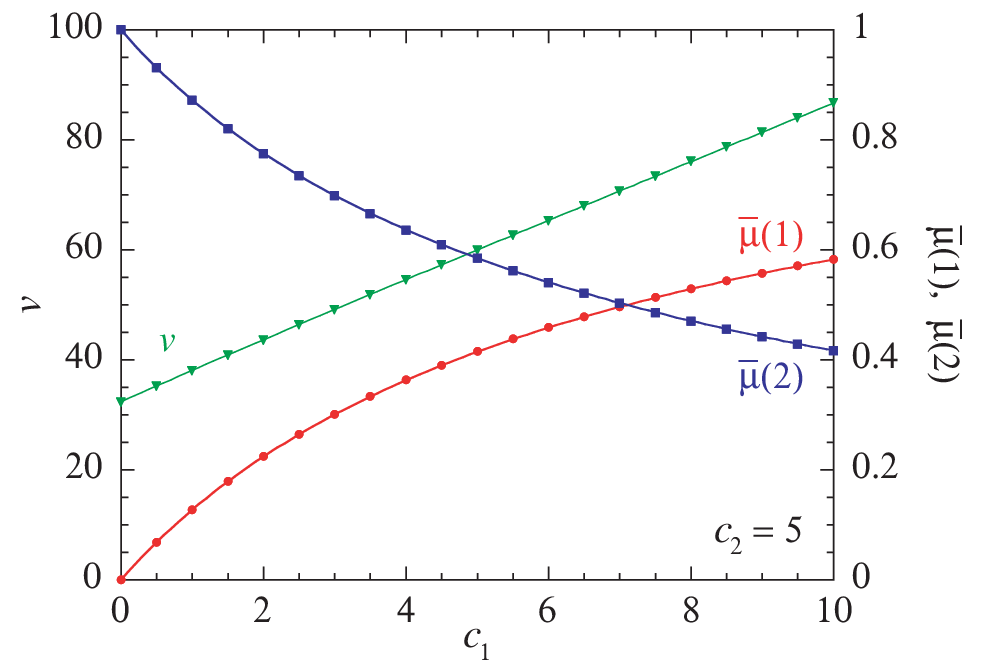}
\caption{The mean growth velocity and the mean fractions $\bar\mu(1)$ and $\bar\mu(2)$ of monomeric units in the grown copy versus the monomer concentration $c_1$ for the fixed value $c_2=5$ of the other monomer concentration.  The template is a Bernoulli sequence of probabilities $\nu(1)=\nu(2)=0.5$ and length $L=10^3$.  The symbols are the numerical results of Gillespie's algorithm with $N_{\rm stat}=10^4$ and the solid lines the theoretical predictions of the iterative method with $N_{\rm loop}=10^2$.}
\label{fig3}
\end{figure}

Figure~\ref{fig3} presents the same quantities as in Fig.~\ref{fig2}, but now as a function of the concentration $c_1$ of monomers of type $m=1$ for a fixed value of the other concentration $c_2=5$.  As the concentration $c_1$ decreases to zero, the fraction $\bar\mu(1)$ of the corresponding monomeric unit $m=1$ in the copy also becomes equal to zero, while $\bar\mu(2)=1-\bar\mu(1)$ reaches the unit value. In the limit $c_1=0$, the copy sequence is thus exclusively composed of units $m=2$, as expected.
Again, we observe the good agreement between the numerical results of Gillespie's algorithm (depicted with dots) and the theoretical predictions of the iterative method (the solid lines).

\begin{figure}[h]
\includegraphics[width=9.5cm]{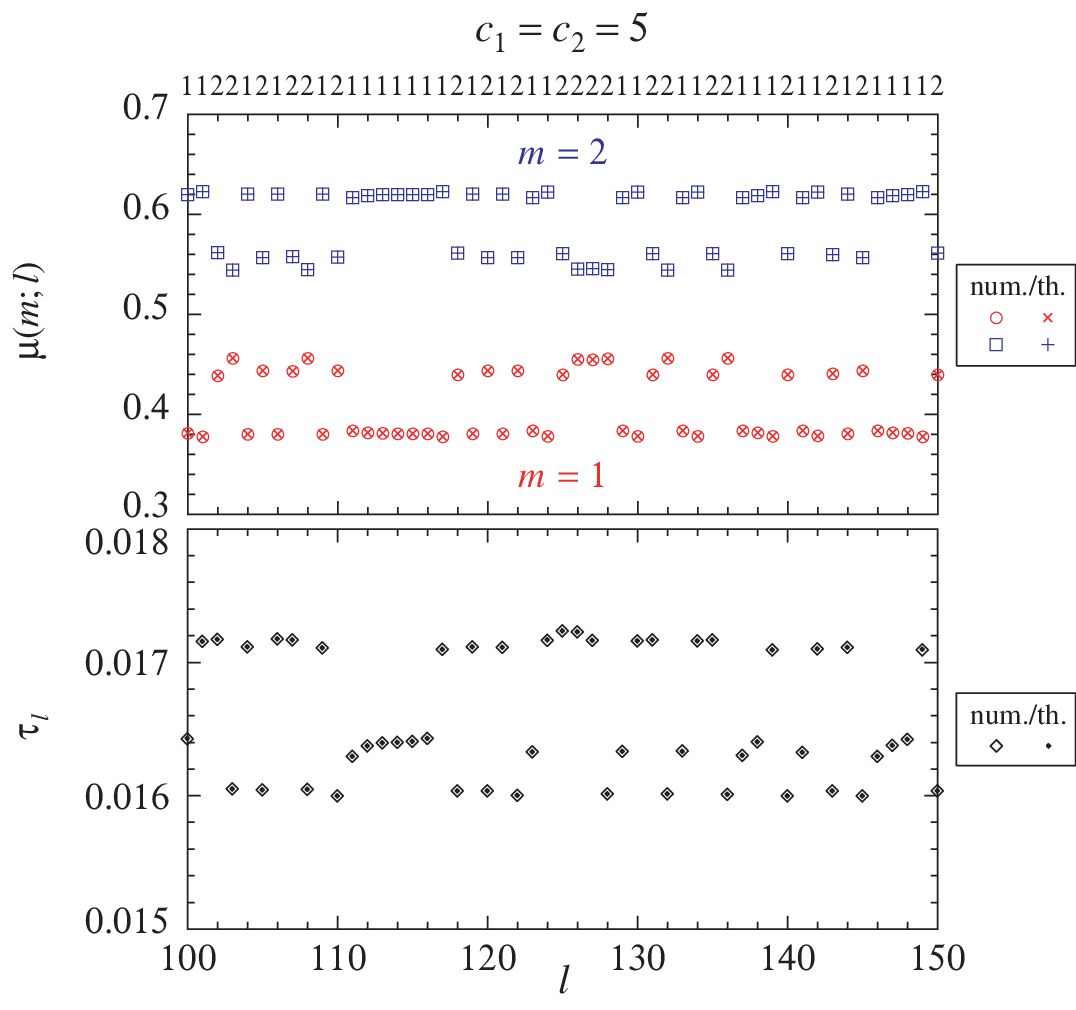}
\caption{The local probabilities $\mu(m;l)$ of monomeric units $m=1$ and $m=2$ (upper panel) and the mean local dwell times $\tau_l$ (lower panel) versus the location $l$ along the template sequence shown above the panels, for the monomer concentrations $c_1=c_2=5$.  The template is a Bernoulli sequence of probabilities $\nu(1)=\nu(2)=0.5$ and length $L=10^3$.  The numerical results of Gillespie's algorithm with $N_{\rm stat}=10^9$ are depicted as open symbols (open circles and open squares for the local monomeric probabilities, open diamonds for the mean local dwell times) and the theoretical predictions of the iterative method with $N_{\rm loop}=10^2$ as crosses and pluses for the local monomeric probabilities and filled diamonds for the mean local dwell times.  The iterative method with $N_{\rm loop}=10^2$ is $4\times 10^6$ times faster than the numerical simulation using Gillespie's algorithm with $N_{\rm stat}=10^9$.}
\label{fig4}
\end{figure}

\begin{figure}[h]
\includegraphics[width=9.5cm]{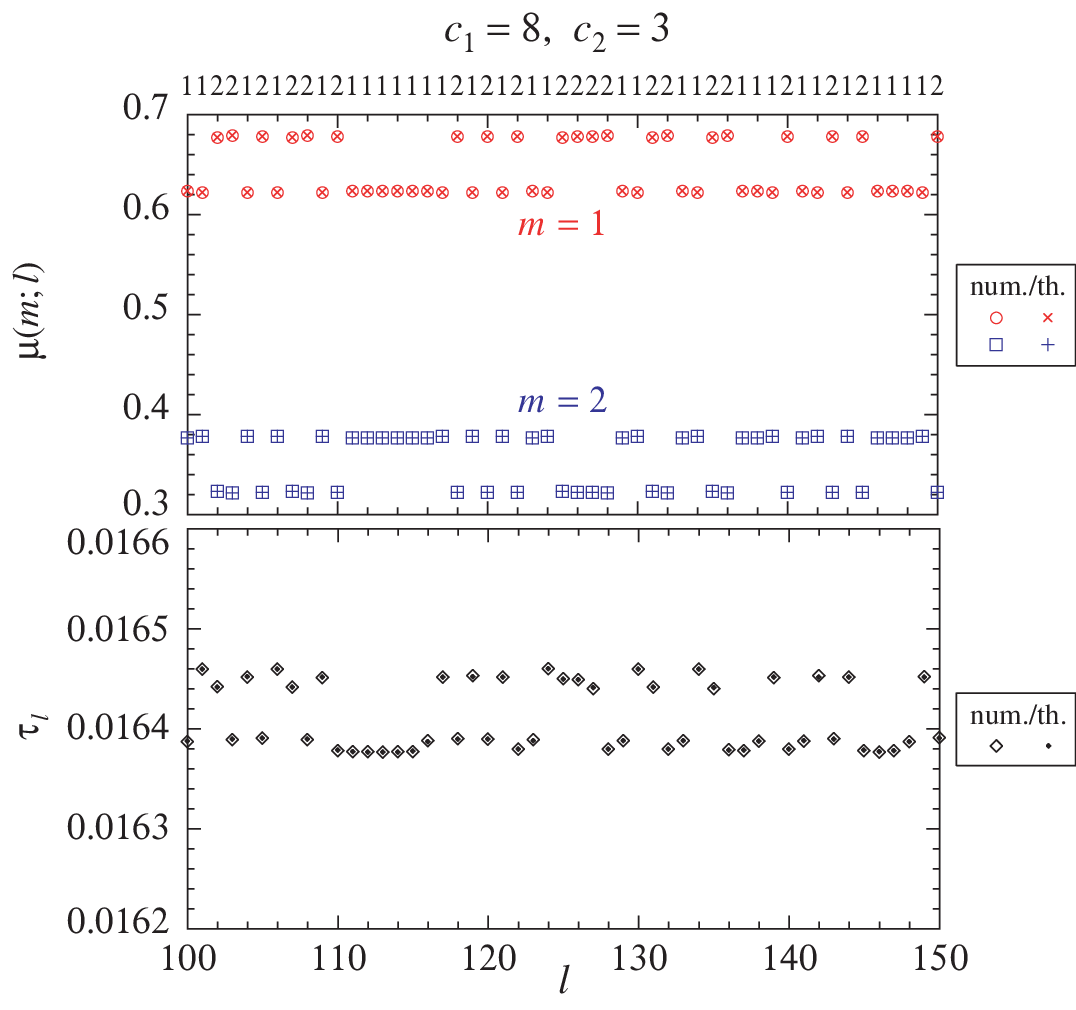}
\caption{The local probabilities $\mu(m;l)$ of monomeric units $m=1$ and $m=2$ (upper panel) and the mean local dwell times $\tau_l$ (lower panel) versus the location $l$ along the template sequence shown above the panels, for the monomer concentrations $c_1=8$ and $c_2=3$.  The template is a Bernoulli sequence of probabilities $\nu(1)=\nu(2)=0.5$ and length $L=10^3$.  The numerical results of Gillespie's algorithm with $N_{\rm stat}=10^9$ are depicted as open symbols (open circles and open squares for the local monomeric probabilities, open diamonds for the mean local dwell times) and the theoretical predictions of the iterative method with $N_{\rm loop}=10^2$ as crosses and pluses for the local monomeric probabilities and filled diamonds for the mean local dwell times.  The iterative method with $N_{\rm loop}=10^2$ is $4\times 10^6$ times faster than the numerical simulation using Gillespie's algorithm with $N_{\rm stat}=10^9$.}
\label{fig5}
\end{figure}

Finally, Figs.~\ref{fig4} and~\ref{fig5} show the local probabilities of monomeric units $\mu(m;l)$ defined by Eq.~(\ref{mu(m;l)}) and the mean local dwell times $\tau_l$ of the catalyst as functions of the location $l$ along the template for two different sets of monomer concentrations.  The open symbols depict the results of Gillespie's algorithm for a statistics over $N_{\rm stat}=10^9$ simulated copies.  The theoretical predictions of the iterative method are plotted as crosses, pluses and filled diamonds.

In Fig.~\ref{fig4}, the concentrations of both monomers are equal to each other, $c_1=c_2=5$.  For these concentrations and the model of rate constants~(\ref{attach_rate_csts})-(\ref{trans_rate_csts}), the copy sequence contains more monomeric units of type $m=2$ than of type $m=1$, on average.  We observe clear correlations between the compositions of the copy and the template, the latter being given above the panels of the figure.  We also observe particular structures in the sequences of local monomeric probabilities $\mu(m;l)$ and mean local dwell times $\tau_l$, when there is a long subsequence of identical units in the template such as the subsequence $1111111$ from $l=111$ to $l=117$.  Contrary to the simplest expectation, the local monomeric probabilities and the mean local dwell times are not uniform along the subsequence, but presents some kind of relaxation toward uniformity at the beginning and the end of the subsequence, after and before switching from and to the other type $m=2$ of monomeric units.  This effect is explained by the fact that the local monomeric probabilities and the mean local dwell times are determined by the backward and forward iterations, running along the template and generating such structures in the successive values of the different quantities of interest.

In Fig.~\ref{fig5}, the concentration $c_1=8$ of monomer $m=1$ is taken to be larger than the concentration $c_2=3$ of the other monomer $m=2$, whereupon the local probabilities $\mu(m;l)$ are now larger for the monomeric units $m=1$ than for the units $m=2$.  In this case, wee see that the mean local dwell times $\tau_l$ vary over a smaller interval of values than in Fig.~\ref{fig4}.  Structures similar as those in Fig.~\ref{fig4} are here also observed, but with much smaller amplitudes.

There is excellent agreement between the results of both methods.  The standard deviation between the numerical and theoretical values can be defined as
\be
\Delta\mu \equiv \left\{ \frac{1}{L_2-L_1} \sum_{l=L_1}^{L_2} \left[ \mu_{\rm num}(m;l)-\mu_{\rm th}(m;l)\right]^2 \right\}^{1/2}
\label{dfn-std}
\ee
for either $m=1$ or $m=2$, since $\mu(1;l)+\mu(2;l)=1$.  The standard deviation~(\ref{dfn-std}) with $L_1=100$ and $L_2=200$  decreases as
\be
\Delta\mu\simeq \frac{0.5 \pm 0.1}{\sqrt{N_{\rm stat}}} \, ,
\ee
as the statistics used in the numerical simulation with Gillespie's algorithm is increased from $N_{\rm stat}=10^3$ to $N_{\rm stat}=10^9$.  But, the point is that the theoretical values of the iterative method are obtained very much faster with only $N_{\rm loop}=10^2$.  Indeed, for the computation of the results plotted in Figs.~\ref{fig4} and~\ref{fig5}, the iterative method is about $4\times 10^6$ times faster than the numerical simulation with Gillespie's algorithm, demonstrating that the iterative method is extremely powerful in providing the exact asymptotic solution of the kinetic equations.

\section{Conclusion and perspectives}
\label{sec:Conclusion}

In this paper, we developed the kinetic theory of the molecular machines performing DNA replication, transcription, and translation, which are the DNA and RNA polymerases and the ribosomes.  In general, these machines have multiple states of conformation or activation and, thus, they catalyze processes of template-directed multistate copolymerization.

Assuming that the rates depend on the internal states of the machine, on the location of the machine along the template, and on the monomeric unit that is attached to or detached from the copy, we show that the kinetic equations of such processes can be solved in the long-time limit to obtain exact formulas for the mean growth velocity, the probabilities of possible sequences of monomeric units composing the copy, and related probability distributions.  Since the rates are assumed to be independent of monomeric units that have been previously incorporated in the copy, the kinetic equations can be exactly solved using the matrix factorization ansatz given by Eq.~(\ref{factorization}) in terms of the matrices~(\ref{Y_ml}), which are defined for every type of monomeric unit possibly found at a given location of the template.  Consequently, the kinetic equations can be solved by running the iteration~(\ref{backward_iter}) backward along the template.  This backward iteration forms an iterated matrix function system (IMFS), each iteration depending on the template monomeric unit (or the group of template units) at the given location.  The backward iteration is complemented by the forward iteration~(\ref{forward_iter}) in order to obtain the mean growth velocity, which is exactly given by the formula~(\ref{v-formula}) with the constant~(\ref{Constant}).  Furthermore, the sequence probabilities are provided by Eq.~(\ref{mu(seq)-formula}) and the local probabilities of monomeric units by Eq.~(\ref{mu(m;l)-formula}).  The detailed mathematical proof of these results is carried out in Appendix~\ref{AppA}.  The IMFS forming the backward iteration can be solved with efficient methods of numerical computation along the template.

The mean growth velocity is positive in the regime of steady growth, where the mean length of the copy increases linearly with time.  If the template sequence is semi-infinite and disordered, there is the possibility for a sublinear growth in time of the copy, as explained in Appendix~\ref{AppB}.  If the attachment rates are too small with respect to the detachment rates, the growth of the copy is stalled.  In the latter regime, the stationary probability distribution of the length of the copy can also be calculated using the backward and forward iterations.

The IMFS method is applied to a numerical example and its results are compared to those obtained by numerical simulations with Gillespie's algorithm.  The latter also gives exact results if the statistics is large enough.  The comparison confirms that both methods provide exact values for the mean growth velocity and the local probabilities of monomeric units for the copy grown along a given template.  In the stationary regime, good agreement is found for the exponent controlling the exponential decrease of the stationary length distribution (see Fig.~\ref{fig1}).  In the growth regime, there is also good agreement between the results of both methods for the mean growth velocity and the fractions of monomeric units in the copy (see Figs.~\ref{fig2} and~\ref{fig3}), and an excellent agreement for the local probabilities of monomeric units in the copy and the mean local dwell times of the catalyst along a given template sequence, as shown in Figs.~\ref{fig4} and~\ref{fig5}.  To compute the results depicted in these figures, the IMFS method is more than a million times faster than the numerical simulation with Gillespie's algorithm. Accordingly, the IMFS method presented in this paper not only provides the exactly asymptotic solution of the kinetic equations, but is also computationally much more efficient than other methods.

In the companion paper \cite{paperII}, the IMFS method will be applied to the kinetics of the T7 DNA polymerase, which exhibits transitions between several conformational states during DNA replication, as the experimental observations show \cite{TJ06,J10,DJ20,DJ21,DKJ22}.

%%%%%%%%%%%%%%%%%%%%%%%%%%%%%%%%%%%%%%%%%%%%%%%%%%%

\begin{acknowledgments}
The author thanks the Universit\'e libre de Bruxelles (ULB) for support.
\end{acknowledgments}

%%%%%%%%%%%%%%%%%%%%%%%%%%%%%%%%%%%%%%%%%%%%%%%%%%%
\appendix

\section{Solving the kinetic equations}
\label{AppA}

\subsection{Equations for the probability matrices~(\ref{P(l)-dfn}), (\ref{P(m,l)-dfn}), (\ref{P(mm',l)-dfn}), $\dots$}

Summing the matrix kinetic equations~(\ref{matrix_kin_eq}) over $m_1\cdots m_l$, we first deduce the equations for the probability matrices~(\ref{P(l)-dfn}):
\bea
\frac{\dd}{\dd t}\, \boldsymbol{\mathsf P}_t(l) &=& \sum_{m_l} \boldsymbol{\mathsf W}_{+m_l,l}^{\rm c} \cdot \boldsymbol{\mathsf P}_t(l-1)
 + \sum_{m_{l+1}} \boldsymbol{\mathsf W}_{-m_{l+1},l+1}^{\rm c} \cdot  \boldsymbol{\mathsf P}_t(m_{l+1} ,l+1)
\nonumber\\
&&+ \, \sum_{m_l}  \Big(\boldsymbol{\mathsf W}_{m_l,l}^{0} - \boldsymbol{\mathsf W}_{-m_l,l}^{\rm d} \Big) \cdot\boldsymbol{\mathsf P}_t(m_l ,l)
- \sum_{m_{l+1}} \boldsymbol{\mathsf W}_{+m_{l+1},l+1}^{\rm d} \cdot\boldsymbol{\mathsf P}_t(l) \, .
\label{matrix_kin_eq_P(l)}
\eea
Next, summing the matrix kinetic equations~(\ref{matrix_kin_eq}) over $m_1\cdots m_{l-1}$, we get the equations for the probability matrices~(\ref{P(m,l)-dfn}):
\bea
\frac{\dd}{\dd t}\, \boldsymbol{\mathsf P}_t(m_l,l) &=& \boldsymbol{\mathsf W}_{+m_l,l}^{\rm c} \cdot \boldsymbol{\mathsf P}_t(l-1) + \sum_{m_{l+1}} \boldsymbol{\mathsf W}_{-m_{l+1},l+1}^{\rm c} \cdot  \boldsymbol{\mathsf P}_t(m_l m_{l+1} ,l+1)
\nonumber\\
&&+ \, \bigg(\boldsymbol{\mathsf W}_{m_l,l}^{0} - \boldsymbol{\mathsf W}_{-m_l,l}^{\rm d} - \sum_{m_{l+1}} \boldsymbol{\mathsf W}_{+m_{l+1},l+1}^{\rm d} \bigg) \cdot\boldsymbol{\mathsf P}_t(m_l ,l)\, .
\label{matrix_kin_eq_P(m,l)}
\eea
Similarly, further equations with the same structure are found for the probability matrices~(\ref{P(mm',l)-dfn}), $\dots$

\subsection{Equations for the particular solutions~(\ref{q-modes})}

Inserting the expressions~(\ref{q-modes}) for the particular solution into Eqs.~(\ref{matrix_kin_eq_P(l)}), (\ref{matrix_kin_eq_P(m,l)}), $\dots$, we infer the following matrix equations for the modes of wave number $q$:
\bea
s_q\, \boldsymbol{\mathsf G}_q(l) &=& {\rm e}^{-\imath q} \sum_{m_l} \boldsymbol{\mathsf W}_{+m_l,l}^{\rm c} \cdot \boldsymbol{\mathsf G}_q(l-1)
 + {\rm e}^{+\imath q}\sum_{m_{l+1}} \boldsymbol{\mathsf W}_{-m_{l+1},l+1}^{\rm c} \cdot  \boldsymbol{\mathsf G}_q(m_{l+1} ,l+1)
\nonumber\\
&&+ \, \sum_{m_l}  \Big(\boldsymbol{\mathsf W}_{m_l,l}^{0} - \boldsymbol{\mathsf W}_{-m_l,l}^{\rm d}\Big) \cdot\boldsymbol{\mathsf G}_q(m_l ,l)
- \sum_{m_{l+1}} \boldsymbol{\mathsf W}_{+m_{l+1},l+1}^{\rm d} \cdot\boldsymbol{\mathsf G}_q(l)\, ,
\label{matrix_kin_eq_G_q(l)}
\eea
\bea
s_q\, \boldsymbol{\mathsf G}_q(m_l,l) &=& {\rm e}^{-\imath q} \, \boldsymbol{\mathsf W}_{+m_l,l}^{\rm c} \cdot \boldsymbol{\mathsf G}_q(l-1) + {\rm e}^{+\imath q} \sum_{m_{l+1}} \boldsymbol{\mathsf W}_{-m_{l+1},l+1}^{\rm c} \cdot  \boldsymbol{\mathsf G}_q(m_l m_{l+1} ,l+1)
\nonumber\\
&&+ \, \bigg(\boldsymbol{\mathsf W}_{m_l,l}^{0} - \boldsymbol{\mathsf W}_{-m_l,l}^{\rm d} - \sum_{m_{l+1}} \boldsymbol{\mathsf W}_{+m_{l+1},l+1}^{\rm d} \bigg) \cdot\boldsymbol{\mathsf G}_q(m_l ,l)\, ,
\label{matrix_kin_eq_G_q(m,l)}
\eea
and so on and so forth.

\subsection{Proof of the matrix factorization~(\ref{factorization})}

Using the expansions~(\ref{s_q}) for $s_q$ and~(\ref{G_q}) for $\boldsymbol{\mathsf G}_q$, and taking the limit $q\to 0$, Eqs.~(\ref{matrix_kin_eq_G_q(l)}), (\ref{matrix_kin_eq_G_q(m,l)}), $\dots$ yield the following equations for the quantities~(\ref{Psi-dfn}):
\be
0 = \sum_{m_l} \boldsymbol{\mathsf W}_{+m_l,l}^{\rm c} \cdot \boldsymbol{\Psi}_{l-1}
 + \sum_{m_{l+1}} \boldsymbol{\mathsf W}_{-m_{l+1},l+1}^{\rm c} \cdot  \boldsymbol{\Psi}_{l+1}(m_{l+1})
+ \sum_{m_l}  \Big(\boldsymbol{\mathsf W}_{m_l,l}^{0} - \boldsymbol{\mathsf W}_{-m_l,l}^{\rm d}\Big) \cdot\boldsymbol{\Psi}_l(m_l )
- \sum_{m_{l+1}} \boldsymbol{\mathsf W}_{+m_{l+1},l+1}^{\rm d} \cdot\boldsymbol{\Psi}_l \, ,
\label{matrix_kin_eq_Psi(l)}
\ee
\be
0 = \boldsymbol{\mathsf W}_{+m_l,l}^{\rm c} \cdot \boldsymbol{\Psi}_{l-1} + \sum_{m_{l+1}} \boldsymbol{\mathsf W}_{-m_{l+1},l+1}^{\rm c} \cdot  \boldsymbol{\Psi}_{l+1}(m_l m_{l+1})
+ \bigg(\boldsymbol{\mathsf W}_{m_l,l}^{0} - \boldsymbol{\mathsf W}_{-m_l,l}^{\rm d} - \sum_{m_{l+1}} \boldsymbol{\mathsf W}_{+m_{l+1},l+1}^{\rm d} \bigg) \cdot\boldsymbol{\Psi}_l(m_l) \, ,
\label{matrix_kin_eq_Psi(m,l)}
\ee
etc. From Eq.~(\ref{matrix_kin_eq_Psi(m,l)}) on, all these matrix equations have a similar structure, so that they can be solved simultaneously using the factorization ansatz~(\ref{factorization}).  Substituting this factorization into Eq.~(\ref{matrix_kin_eq_Psi(m,l)}) and supposing it holds for any $\boldsymbol{\Psi}_{l-1}$, we find the following matrix equation for $\boldsymbol{\mathsf Y}_{m_l,l}$:
\be
0 = \boldsymbol{\mathsf W}_{+m_l,l}^{\rm c} + \bigg( \sum_{m_{l+1}} \boldsymbol{\mathsf W}_{-m_{l+1},l+1}^{\rm c} \cdot  \boldsymbol{\mathsf Y}_{m_{l+1},l+1} + \boldsymbol{\mathsf W}_{m_l,l}^{0} - \boldsymbol{\mathsf W}_{-m_l,l}^{\rm d} - \sum_{m_{l+1}} \boldsymbol{\mathsf W}_{+m_{l+1},l+1}^{\rm d} \bigg) \cdot\boldsymbol{\mathsf Y}_{m_l,l} \, .
\label{matrix_kin_eq_Y(m,l)}
\ee
Consequently, all the equations of the hierarchy beyond Eq.~(\ref{matrix_kin_eq_Psi(l)}) are solved as soon as Eq.~(\ref{matrix_kin_eq_Y(m,l)}) is satisfied.

\subsection{Proof of the backward iteration~(\ref{backward_iter})}

Now, if we define the following matrix,
\be
\boldsymbol{\mathsf V}_{l} \equiv \sum_{m_{l+1}} \boldsymbol{\mathsf W}_{+m_{l+1},l+1}^{\rm d} - 
\sum_{m_{l+1}} \boldsymbol{\mathsf W}_{-m_{l+1},l+1}^{\rm c} \cdot  \boldsymbol{\mathsf Y}_{m_{l+1},l+1} \, ,
\label{V_l-dfn}
\ee
and use it inside the parenthesis of~Eq.~(\ref{matrix_kin_eq_Y(m,l)}), the expression~(\ref{Y_ml}) for the matrix $\boldsymbol{\mathsf Y}_{m_l,l}$ is obtained after the inversion of Eq.~(\ref{matrix_kin_eq_Y(m,l)}).  Replacing $l$ with $l-1$ in Eq.~(\ref{V_l-dfn}) and substituting therein the expression~(\ref{Y_ml}), we find the backward iteration~(\ref{backward_iter}).

Therefore, all the equations of the hierarchy, starting from Eq.~(\ref{matrix_kin_eq_Psi(m,l)}) and going on for the quantities~(\ref{factorization}) are solved once the backward iteration~(\ref{backward_iter}) has computed the solutions for all the matrices~$\boldsymbol{\mathsf V}_{l}$ and $\boldsymbol{\mathsf Y}_{m_l,l}$ along the template.

\subsection{Proof of Eq.~(\ref{v-formula}) for the mean growth velocity}

If we take the derivative of Eq.~(\ref{matrix_kin_eq_G_q(l)}) with respect to the wave number $q$ and set $q=0$, we get
\bea
-\imath \, v \, \boldsymbol{\Psi}_l &=& \sum_{m_l} \boldsymbol{\mathsf W}_{+m_l,l}^{\rm c} \cdot \big( -\imath\,  \boldsymbol{\Psi}_{l-1}+ \boldsymbol{\Psi}_{l-1}^{\prime}\big)
 + \sum_{m_{l+1}} \boldsymbol{\mathsf W}_{-m_{l+1},l+1}^{\rm c} \cdot \Big[\imath \, \boldsymbol{\Psi}_{l+1}(m_{l+1}) + \boldsymbol{\Psi}_{l+1}^{\prime}(m_{l+1}) \Big] \nonumber\\
 && 
+ \sum_{m_l}  \Big(\boldsymbol{\mathsf W}_{m_l,l}^{0} - \boldsymbol{\mathsf W}_{-m_l,l}^{\rm d}\Big) \cdot \boldsymbol{\Psi}_l^{\prime}(m_l)
- \sum_{m_{l+1}} \boldsymbol{\mathsf W}_{+m_{l+1},l+1}^{\rm d} \cdot\boldsymbol{\Psi}_l^{\prime}
\label{matrix_kin_eq_Psi^prime(l)}
\eea
in terms of the quantities~(\ref{Psi-dfn}) and~(\ref{Psi'-dfn}).

In order to deduce an expression for the mean growth velocity $v$, we first take the trace of Eq.~(\ref{matrix_kin_eq_Psi^prime(l)}) and use the following identities satisfied by the matrices~(\ref{Wc-matrix}), (\ref{Wd-matrix}), and~(\ref{W0-matrix}):
\be
{\rm tr}(\, \boldsymbol{\mathsf W}_{m,l}^{0}  \cdot \boldsymbol{\mathsf X}) = 0
\label{tr-W0-X=0}
\ee
and
\be
{\rm tr}(\,\boldsymbol{\mathsf W}_{\pm m,l}^{\rm c}  \cdot \boldsymbol{\mathsf X}) = {\rm tr}(\,\boldsymbol{\mathsf W}_{\pm m,l}^{\rm d}  \cdot \boldsymbol{\mathsf X})
\label{tr-Wc-X=tr-Wd-X}
\ee
for any $I\times I$ matrix of the form
\be
\boldsymbol{\mathsf X} =
\left[
\begin{array}{cccc}
x^1 & x^1 & \dots & x^1 \\
x^2 & x^2 & \dots & x^2 \\
\vdots & \vdots & \ddots & \vdots \\
x^I & x^I & \dots & x^I
\end{array}
\right] ,
\label{X-matrix}
\ee
having $I$ identical columns of size $I$.

Next, we sum the trace of Eq.~(\ref{matrix_kin_eq_Psi^prime(l)}) over the period $L$ of the template, which allows us to replace the subscripts $l+1$ with $l$.  Because the identities~(\ref{tr-W0-X=0}) and~(\ref{tr-Wc-X=tr-Wd-X}) hold with $\boldsymbol{\mathsf X}=\boldsymbol{\Psi}_l^{\prime}(\cdot)$, all the corresponding terms cancel and we find
\be
v \, \sum_{l=1}^L {\rm tr}\, \boldsymbol{\Psi}_l = \sum_{l=1}^L \sum_{m_l} {\rm tr}\big(\boldsymbol{\mathsf W}_{+m_l,l}^{\rm c} \cdot \boldsymbol{\Psi}_{l-1}\big) - \sum_{l=1}^L \sum_{m_l} {\rm tr}\big[\boldsymbol{\mathsf W}_{-m_l,l}^{\rm c} \cdot \boldsymbol{\Psi}_l(m_l) \big] \, .
\label{eq-v}
\ee
If the identity~(\ref{tr-Wc-X=tr-Wd-X}) for $\boldsymbol{\mathsf X}=\boldsymbol{\Psi}_{l-1}$ is used together with Eq.~(\ref{V_l-dfn}) after the substitution $l\to l-1$ and with Eq.~(\ref{Psi_ml-Y_ml}), we find that the mean growth velocity is given by
\be
v = \frac{\sum_{l=1}^L {\rm tr}\big(\boldsymbol{\mathsf V}_{l-1}\cdot\boldsymbol{\Psi}_{l-1}\big)}{\sum_{l=1}^L {\rm tr}\, \boldsymbol{\Psi}_l} \, .
\label{eq-v-2}
\ee
Subsequently, the formula~(\ref{v-formula}) for the mean growth velocity is obtained in the limit $L\to\infty$, if the invariance property~(\ref{Constant}) holds, which is proved in the next Subsection~\ref{AppA:property_C}.

\subsection{Proof of the invariance property~(\ref{Constant})}
\label{AppA:property_C}

Now, Eq.~(\ref{Constant}) can be proved as follows.  Using again Eq.~(\ref{V_l-dfn}) with $l\to l-1$, we deduce
\bea
{\rm tr}(\boldsymbol{\mathsf V}_{l-1} \cdot \boldsymbol{\Psi}_{l-1}) &=& \sum_{m_l} {\rm tr}\big[ ( \boldsymbol{\mathsf W}_{+m_l,l}^{\rm d} - \boldsymbol{\mathsf W}_{-m_l,l}^{\rm c} \cdot  \boldsymbol{\mathsf Y}_{m_l,l}) \cdot \boldsymbol{\Psi}_{l-1}\big]
\nonumber\\
&=& \sum_{m_l} {\rm tr}\big[ ( \boldsymbol{\mathsf W}_{+m_l,l}^{\rm c} - \boldsymbol{\mathsf W}_{-m_l,l}^{\rm c} \cdot  \boldsymbol{\mathsf Y}_{m_l,l}) \cdot \boldsymbol{\Psi}_{l-1}\big]
\nonumber\\
&=& \sum_{m_l} {\rm tr}\big[ ( \boldsymbol{\mathsf V}_l - \boldsymbol{\mathsf W}_{m_l,l}^{0} + \boldsymbol{\mathsf W}_{-m_l,l}^{\rm d} - \boldsymbol{\mathsf W}_{-m_l,l}^{\rm c} ) \cdot \boldsymbol{\Psi}_l(m_l)\big]
\nonumber\\
&=& {\rm tr}(\boldsymbol{\mathsf V}_l \cdot \boldsymbol{\Psi}_l) \, .
\label{proof-Constant}
\eea
From the first to the second line, we used the identity~(\ref{tr-Wc-X=tr-Wd-X}) for $\boldsymbol{\mathsf X}=\boldsymbol{\Psi}_{l-1}$; from the second to the third line, the expression~(\ref{Y_ml}) for $\boldsymbol{\mathsf Y}_{m_l,l}$ and the relation~(\ref{Psi_ml-Y_ml}); and, finally, from the third to the fourth line, the identities~(\ref{tr-W0-X=0}) and~(\ref{tr-Wc-X=tr-Wd-X}) for $\boldsymbol{\mathsf X}=\boldsymbol{\Psi}_l(m_l)$.  Thus, the invariance property~(\ref{Constant}) is proved.

\section{Time evolution of the length distribution}
\label{AppB}

\subsection{Equation for the length distribution}

The equation for the length distribution~(\ref{p(l)-trace}) can be deduced by taking the trace of Eq.~(\ref{matrix_kin_eq_P(l)}) and using the identities~(\ref{tr-W0-X=0}) and~(\ref{tr-Wc-X=tr-Wd-X}) for $\boldsymbol{\mathsf X}=\boldsymbol{\mathsf P}_t(\cdot)$.  We find that
\be
\frac{\dd}{\dd t}\, p_t(l) = a_l \, p_t(l-1) + b_{l+1} \, p_t(l+1) - (a_{l+1}+b_l) \, p_t(l)
\label{eq-p_t(l)}
\ee
with the following coefficients,
\bea
a_l &\equiv& \frac{1}{{\rm tr}\,\boldsymbol{\mathsf P}_t(l-1)} \, \sum_{m_l} {\rm tr}\big[\boldsymbol{\mathsf W}_{+m_l,l}^{\rm c} \cdot \boldsymbol{\mathsf P}_t(l-1)\big] \, , \label{a_l-dfn} \\
b_l &\equiv& \frac{1}{{\rm tr}\,\boldsymbol{\mathsf P}_t(l)} \, \sum_{m_l} {\rm tr}\big[\boldsymbol{\mathsf W}_{-m_l,l}^{\rm c} \cdot \boldsymbol{\mathsf P}_t(m_l,l)\big] \, . \label{b_l-dfn}
\eea
In the long-time limit, the sequence probabilities evolve toward stationary distributions given by the expressions~(\ref{factorization}), so that these coefficients can be computed according to
\bea
a_l &=& \frac{1}{{\rm tr}\,\boldsymbol{\Psi}_{l-1}} \, \sum_{m_l} {\rm tr}\big(\boldsymbol{\mathsf W}_{+m_l,l}^{\rm c} \cdot \boldsymbol{\Psi}_{l-1}\big) \, , \label{a_l-eq} \\
b_l &=& \frac{1}{{\rm tr}\,\boldsymbol{\Psi}_l} \, \sum_{m_l} {\rm tr}\big(\boldsymbol{\mathsf W}_{-m_l,l}^{\rm c} \cdot \boldsymbol{\mathsf Y}_{m_l,l}\cdot \boldsymbol{\Psi}_{l-1}\big) \, , \label{b_l-eq}
\eea
in terms of the matrices $\boldsymbol{\Psi}_l$ given by the forward iteration~(\ref{forward_iter}) and the matrices $\boldsymbol{\mathsf Y}_{m_l,l}$ defined by Eq.~(\ref{Y_ml}) and calculated with the backward iteration~(\ref{backward_iter}).

\subsection{The regime of sublinear growth in time}

If the copolymer chain grows along a disordered template, its mean length may increase sublinearly in time as $\langle l \rangle_t \sim t^\gamma$ with an exponent $0<\gamma<1$, which is the root of the following equation:
\be
\left\langle \left( \frac{b_l}{a_l}\right)^\gamma \right\rangle = 1 \, ,
\label{eq-gamma}
\ee
as shown for random drifts in one-dimensional disordered media~\cite{D83,BG90} and for unistate templated-directed copolymerization processes~\cite{G17JSM}.  Therefore, the exponent $\gamma$ can be computed using Eqs.~(\ref{a_l-eq}) and~(\ref{b_l-eq}) for the coefficients $a_l$ and $b_l$.

\subsection{Application to the numerical example}

The formula~(\ref{eq-gamma}) is applied to the numerical example of rate constants~(\ref{attach_rate_csts})-(\ref{trans_rate_csts}) for a template of length $L=10^4$, finding the root of $\langle(b_l/a_l)^\gamma\rangle=L^{-1}\sum_{l=1}^L(b_l/a_l)^\gamma=1$ as a function of the concentrations.

In the case of Figs.~\ref{fig1} and~\ref{fig2}, where the concentrations are equal, we find that
\bea
&&\gamma = 0 \qquad\mbox{if}\qquad c_1=c_2=0.2974\pm 0.0001 \, , \\
&&\gamma = 1 \qquad\mbox{if}\qquad c_1=c_2=0.2976\pm 0.0001 \, ,
\eea
which confirms the values~(\ref{c1=c2_stat_reg}) and~(\ref{c1=c2_st_growth_reg}) obtained in Section~\ref{sec:num_example}, up to numerical accuracy.

A similar computation can be performed for other ranges of concentrations.  For instance, we find
\bea
\mbox{for}\quad c_2=0.1: \qquad &&\gamma = 0 \qquad\mbox{if}\qquad c_1=0.584\pm 0.005 \, , \\
&&\gamma = 1 \qquad\mbox{if}\qquad c_1=0.615\pm 0.005 \, ,
\eea
and
\bea
\mbox{for}\quad c_2=0.01: \qquad &&\gamma = 0 \qquad\mbox{if}\qquad c_1=0.732\pm 0.005 \, , \\
&&\gamma = 1 \qquad\mbox{if}\qquad c_1=0.792\pm 0.005 \, .
\eea
Accordingly, a broader interval of sublinear growth in time is found if one of the monomer concentrations takes a relatively small fixed value.

\section{Numerical simulation with Gillespie's algorithm}
\label{AppC}

The Markov jump stochastic process of reactions~(\ref{kin_1}) and~(\ref{kin_2}) can be exactly simulated with Gillespie's numerical algorithm, which is of Monte Carlo type \cite{G76,G77}.  

The algorithm aims at generating a random trajectory of the process and, thus, a random copy of the template.  At every step of the simulation, we may assume that, before the jump $\omega\to\omega^\prime$, the copy has the length $l$, its ultimate unit is $m_l$, which is paired with the template unit $n_l$, the next template unit is $n_{l+1}$, and the catalyst is found in its internal state $i$.  One of the following seven possible jumps may occur:  1) \& 2) the attachment of either the unit $m_{l+1}=1$ or $m_{l+1}=2$ to the copy at the rate $w_{+m_{l+1},n_{l+1}}^{i\to i}$ with the elongation $l\to l+1$ for the copy; 3) \& 4) the attachment of either the unit $m_{l+1}=1$ or $m_{l+1}=2$ to the copy at the rate  $w_{+m_{l+1},n_{l+1}}^{i\to j}$ with the transition $i\to j\ne i$ for the catalyst, and the elongation $l\to l+1$ for the copy; 5) the detachment of the unit $m_l$ from the copy at the rate $w_{-m_l,n_l}^{i\to i}$, so that $l\to l-1$; 6) the detachment of the unit $m_l$ from the copy at the rate $w_{-m_l,n_l}^{i\to j}$ with the transition $i\to j\ne i$ for the catalyst and $l\to l-1$; 7) the transition $i\to j\ne i$ for the catalyst at the rate $w_{0,n_l}^{i\to j}$ without any attachment or detachment, so that $l\to l$.  For the initiation steps~(\ref{kin0_1})-(\ref{kin0_2}), there are only five possible jumps: 1) \& 2) the attachment of either $m_1=1$ or $m_1=2$ without transition for the catalyst; 3) \& 4) the attachment of either $m_1=1$ or $m_1=2$ with the transition $i\to j\ne i$ for the catalyst; 5) the transition $i\to j\ne i$ without attachment.

The random waiting time $\Delta t(\omega\to\omega^\prime)$ before the jump $\omega\to\omega^\prime$ has the exponential probability density $p(\Delta t)={\mathcal T}^{-1} \exp(-\Delta t/{\mathcal T})$ with the parameter ${\mathcal T}=\Big[\sum_{\omega^\prime} w(\omega\to\omega^\prime)\Big]^{-1}$, given by the inverse of the sum of the rates for the seven possible jumps.  The random jump $\omega\to\omega^\prime$ occurs with the probability ${\mathcal P}(\omega\to\omega^\prime)=w(\omega\to\omega^\prime)/\sum_{\omega^\prime} w(\omega\to\omega^\prime)$, given by the ratio of the corresponding rate over the sum of rates.

The mean local dwell time $\tau_l$ of the catalyst of the template is given by the sum of all the random waiting times $\Delta t(\omega\to\omega^\prime)$ of the jumps occurring at the location $l$ of the template when the copy has its length equal to $l$, and by averaging over a large enough statistical ensemble of $N_{\rm stat}$ random copies generated with Gillespie's algorithm: $\tau_l =\langle\sum_{\omega:l}\Delta t(\omega\to\omega^\prime)\rangle_{N_{\rm stat}}$.

Two uniform pseudorandom numbers are generated at every step: one for the random waiting time and the other for selecting a random jump among the seven possibilities.  This is repeated over $N_{\rm step}$ steps, taking $N_{\rm step}=L$ equal to the length $L$ of the template in order to keep the copy shorter than the template.  The growth velocity of an individual random trajectory is computed as the length of the copy at the end of the $N_{\rm step}$ steps divided by the sum of all the random waiting times generated during the simulation of the trajectory.  The mean growth velocity is then given by averaging over the statistical ensemble of trajectories: $v=\langle l_{N_{\rm step}}/\sum_{N_{\rm step}}\Delta t(\omega\to\omega^\prime)\rangle_{N_{\rm stat}}$.  In addition, the local monomeric probabilities $\mu(m;l)$ are also computed with the statistical ensemble of $N_{\rm stat}$ random copies.

%%%%%%%%%%%%%%%%%%%%%%%%%%%%%%%%%%%%%%%%%%%%%%

%%%%%%%%%%%%%%%%%%%%%%%%%%%%%%%%%%%%%%%%%%%%%%%%%%%

\end{document}